\documentclass[aps,prb,twocolumn,superscriptaddress,floatfix,showpacs,10pt,letterpaper]{revtex4}

\pdfoutput=1

\newcommand{\fH}{\mathscr{H}}

\newcommand{\fZ}{\mathscr{Z}}

\newcommand{\RZ}{\R/\Z}

\begin{document}

\begin{titlepage}

\title{Classifying gauge anomalies through SPT orders\\
and classifying gravitational anomalies through topological orders
}

\author{Xiao-Gang Wen} 
\affiliation{Perimeter Institute for Theoretical Physics, Waterloo, Ontario, N2L 2Y5 Canada} 
\affiliation{Department of Physics, Massachusetts Institute of Technology, Cambridge, Massachusetts 02139, USA}

\begin{abstract} 
In this paper, we systematically  study gauge anomalies in bosonic and
fermionic weak-coupling gauge theories with gauge group $G$ (which can be
continuous or discrete) in $d$ space-time dimensions.  We show a very close
relation between gauge anomalies for gauge group $G$ and symmetry-protected
trivial (SPT) orders [also known as symmetry-protected topological (SPT)
orders] with symmetry group $G$ in one higher dimension. The SPT phases are
classified by group cohomology class $\cH^{d+1}(G,\RZ)$.  
Through a more careful consideration, we argue that, the gauge anomalies are
described by the elements in Free$[\cH^{d+1}(G,\R/\Z)]\oplus
\fH_\pi^{d+1}(BG,\R/\Z)$.  The well known Adler-Bell-Jackiw anomalies are
classified by the free part of $\cH^{d+1}(G,\R/\Z)$ (denoted as
Free$[\cH^{d+1}(G,\R/\Z)]$).  We refer other kinds of gauge anomalies beyond
Adler-Bell-Jackiw anomalies as nonABJ gauge anomalies, which include Witten
$SU(2)$ global gauge anomaly.  We introduce a notion of $\pi$-cohomology group,
$\fH_\pi^{d+1}(BG,\R/\Z)$, for the classifying space $BG$, which is an Abelian
group and include Tor$[\cH^{d+1}(G,\R/\Z)]$ and topological cohomology group
$H^{d+1}(BG,\R/\Z)$ as subgroups.  We argue that $\fH_\pi^{d+1}(BG,\R/\Z)$
classifies the bosonic nonABJ gauge anomalies, and partially classifies
fermionic nonABJ anomalies.
Using the same approach that shows  gauge anomalies to be connected to SPT
phases, we can also show that gravitational anomalies are connected to
topological orders (\ie patterns of long-range entanglement) in one-higher
dimension.

\end{abstract}


\maketitle

\end{titlepage}

{\small \setcounter{tocdepth}{1} \tableofcontents }

\section{Introduction}

Gauge anomaly in a gauge theory is a sign that the theory is not well defined.
The first known gauge anomaly is Adler-Bell-Jackiw anomaly.\cite{A6926,BJ6947}
The second type of  gauge anomaly is Witten $SU(2)$ global anomaly.\cite{W8224}
Some recent work on gauge anomaly can be found in
\Ref{CP9045,BC0099,FMS0315,HS0629,GSW1113,FGT1317}.  Those anomalies are for
continuous gauge groups.  The gauge anomalies can also appear for discrete
gauge groups.  Previously, the understanding of those discrete-group anomalies
was obtained by embedding the discrete gauge groups into continuous gauge
groups,\cite{BD9224,CM9814} which only captures part of gauge anomalies for
discrete gauge groups.  

In condensed matter physics, close relations between gauge/gravitational
anomalies and gapless edge excitations\cite{W9125,KF9732} in quantum Hall
states \cite{KDP8094,TSG8259} have being found. Also close relations between
gauge/gravitational anomalies of continuous groups and topological
insulators/superconductors\cite{KM0501,BZ0602,KM0502,MB0706,R0922,FKM0703,QHZ0824,K0986,SRF0825,SMF9945,RG0067,R0664,QHR0901,SF0904}
have been observed,\cite{RML1204,RMO0701,S1295,HLP1242,RS1296} which were used
extensively to understand and study topological
insulators/superconductors.\cite{RML1204} 

In this paper, we will give a systematic understanding of gauge anomalies in
weak-coupling gauge theories, where weakly fluctuating gauge fields are coupled
to matter fields.  If the matter fields are all bosonic, the corresponding gauge
anomalies are called bosonic gauge anomalies.  If some matter fields are
fermionic, the corresponding gauge anomalies are called fermionic gauge
anomalies.  We find that we can gain a systematic understanding of gauge
anomalies through SPT states, which allow us to understand gauge anomalies for
both continuous and discrete gauge groups directly.  

What are SPT states?  SPT states are short-range entangled states with an
on-site symmetry described by the symmetry group $G$.\cite{GW0931,PBT0959}  It
was shown that one can use distinct elements in group cohomology class
$\cH^{d+1}(G,\RZ)$ to construct distinct SPT states in $(d+1)$-dimensional
space-time.\cite{CLW1141,CGL1172,CGL1204}  

The SPT states have very special low energy boundary effective theories, where
the symmetry $G$ in the bulk is realized as a \emph{non-on-site} symmetry on
the boundary.  If we try to gauge the non-on-site symmetry, we will get an
anomalous gauge theory, as demonstrated in
\Ref{CGL1172,LV1219,LW1224,CW1217,SL1204} for $G=U(1),SU(2)$.  This relation
between SPT states and gauge anomalies on the boundary of the SPT states is
called anomaly inflow (the first example was discovered in \Ref{L8132,CH8527}),
which allows us to obtain the following result \frm{one can use different
elements in group cohomology class $\cH^{d+1}(G,\RZ)$ to construct different
bosonic gauge anomalies for gauge group $G$ in $d$-dimensional space-time.}
This result applies for both continuous and discrete gauge groups.  The free
part of $\cH^{d+1}(G,\RZ)$, Free$[\cH^{d+1}(G,\RZ)]$, classifies the well known
Adler-Bell-Jackiw  anomaly for both bosonic and fermionic systems.  The torsion
part of $\cH^{d+1}(G,\RZ)$ correspond to new types of gauge anomalies beyond
the Adler-Bell-Jackiw anomaly (which will be called nonABJ gauge anomalies).  

However, in the above systematic description, the non-trivial gauge anomalies
come from the non-trivial homological structure of the classifying space $BG$
of the gauge group $G$.  On the other hand, we know that non-trivial  global
anomalies come from non-trivial homotopic structure $\pi_d(G)$ of 
$G$, which is the same as the homotopic structure of the classifying space
since $\pi_{d+1}(BG)=\pi_d(G)$.  Therefore, the cohomology description of gauge
anomalies may miss some  global anomalies which can only be captured by the
homotopic structure of $BG$, instead of the homological structure.

In an attempt to obtain a more general description of gauge anomalies, we
introduce a notion of $\pi$-cohomology group $\fH_\pi^{d+1}(BG,\R/\Z)$ for the
classifying space $BG$ of the gauge group $G$.  $\fH_\pi^{d+1}(BG,\R/\Z)$ is an
Abelian group which include the topological cohomology class
$H^{d+1}(BG,\R/\Z)$ and group cohomology class Tor$[\cH^{d+1}(G,\R/\Z)]$ as
subgroups (see appendix \ref{HH}):
\begin{align}
 \text{Tor}[\cH^{d+1}(G,\R/\Z)] \subset H^{d+1}(BG,\R/\Z)
\subset \fH_\pi^{d+1}(BG,\R/\Z).
\end{align}
If $G$ is finite, we further have
\begin{align}
 \text{Tor}[\cH^{d+1}(G,\R/\Z)] = H^{d+1}(BG,\R/\Z)
= \fH_\pi^{d+1}(BG,\R/\Z).
\end{align}
We like to remark that, by definition, $\fH_\pi^{d+1}(BG,\RZ)$ is more general
than $H^{d+1}(BG,\RZ)$. But at the moment, we do not know if
$\fH_\pi^{d+1}(BG,\RZ)$ is strictly larger than $H^{d+1}(BG,\RZ)$.  It is still
possible that $\fH_\pi^{d+1}(BG,\R/\Z) = H^{d+1}(BG,\R/\Z)$ even for
continuous group.

We find that we can use the different elements in the $\pi$-cohomology group
$\fH_\pi^{d+1}(BG,\R/\Z)$ to construct different nonABJ gauge anomalies.
So more generally, \frm{the bosonic/fermionic gauge anomalies are described by
Free$[\cH^{d+1}(G,\RZ)]\oplus \fH_\pi^{d+1}(BG,\R/\Z)$. It is possible that
Free$[\cH^{d+1}(G,\RZ)]\oplus \fH_\pi^{d+1}(BG,\R/\Z)$ classify all the bosonic
gauge anomalies.  $\fH_\pi^{d+1}(BG,\R/\Z)$ includes $H^{d+1}(BG,\RZ)$ as a
subgroup.  }

We note that Witten's $SU(2)$ global anomaly is a fermionic global anomaly with
known realization by fermionic systems.  Since the $\pi$-cohomology result
Free$[\cH^{d+1}(G,\RZ)]\oplus \fH_\pi^{d+1}(BG,\R/\Z)$ only describes part of
fermionic gauge anomalies, it is not clear if it includes Witten's $SU(2)$
global anomaly.  On the other hand, we know for sure that the group cohomology
result $H^{d+1}(BG,\R/\Z)$ does not include the $SU(2)$ global anomaly since
$H^{5}(BSU(2),\R/\Z)=0$.

We will define $\fH_\pi^{d+1}(BG,\R/\Z)$ later in section \ref{more}.  In the
next two sections, we will first give a general picture of our approach and
present some simple examples of the new nonABJ gauge anomalies.  Then we will
give a general systematic discussion of gauge anomalies, and their
description/classification in terms of Free$[\cH^{d+1}(G,\RZ)]\oplus
\fH_\pi^{d+1}(BG,\R/\Z)$.

Last, we will use the connection between gauge anomalies and SPT phases (in one
higher dimension) to construct a non-perturbative definition of any
anomaly-free chiral gauge theories.  We find that even certain anomalous
chiral gauge theories can be defined  non-perturbatively.

\section{A general discussion of gauge anomalies}
\label{gen}

\subsection{Study gauge anomalies in one-higher dimension and in zero-coupling
limit}

We know that anomalous  gauge theories are not well defined. But, how can we
classify something that are not well defined?  We note that if we view a gauge
theory with the Adler-Bell-Jackiw anomaly in $d$-dimensional space-time as the
boundary of a theory in $(d+1)$-dimensional space-time, then the combined
theory is well defined.  The gauge non-invariance of the anomalous boundary
gauge theory is canceled by the gauge non-invariance of a Chern-Simons term on
$(d+1)$-dimensional bulk which is gauge invariant only up to a boundary term.  
So we define $d$-dimensional anomalous  gauge theories through defining a
$(d+1)$-dimensional bulk theory. The classification of the $(d+1)$-dimensional
bulk theories will leads to a classification of anomalies in $d$-dimensional
gauge theories.

The $(d+1)$-dimensional bulk theory has the following generic form
\begin{align}
 \cL_{d+1D}=\cL_{d+1D}^\text{matter}(\phi,\psi,A_\mu) 
+ \frac{\Tr(F_{\mu\nu}F^{\mu\nu})}{\la_g}
\end{align}
where $\phi$ (or $\psi$) are bosonic (or fermionic) matter fields that couple
to a gauge field $A_\mu$ of gauge group $G$.  In this paper, we will study
gauge anomalies in weak-coupling gauge theory.  So we can take the
zero-coupling limit: $\la_g\to 0$.  In this limit we can treat the gauge field
$A_\mu$ as non-dynamical probe field and study only the theory of the mater
fields $\cL_{d+1D}^\text{matter}(\phi,\psi,A_\mu)$, which has an on-site
symmetry with symmetry group $G$ if we set the probe field $A_\mu=0$.  So we
can study  $d$-dimensional gauge anomalies through $(d+1)$-dimensional bulk
theories with only matter and an  on-site symmetry $G$.  

\subsection{Gauge anomalies and SPT states}

Under the above set up, the problem of gauge anomaly becomes the following
problem: \frm{Given a low energy theory with a global symmetry $G$ in
$d$-dimensional space-time, is there a non-perturbatively well-defined theory
with \emph{on-site symmetry} in the same dimension which reproduce the low
energy theory.} We require the  global symmetry $G$ to be an on-site symmetry
in the well-defined theory since we need to gauge the global symmetry to
recover the gauge theory with gauge group $G$.

\begin{figure}[tb] 
\begin{center} 
\includegraphics[scale=0.4]{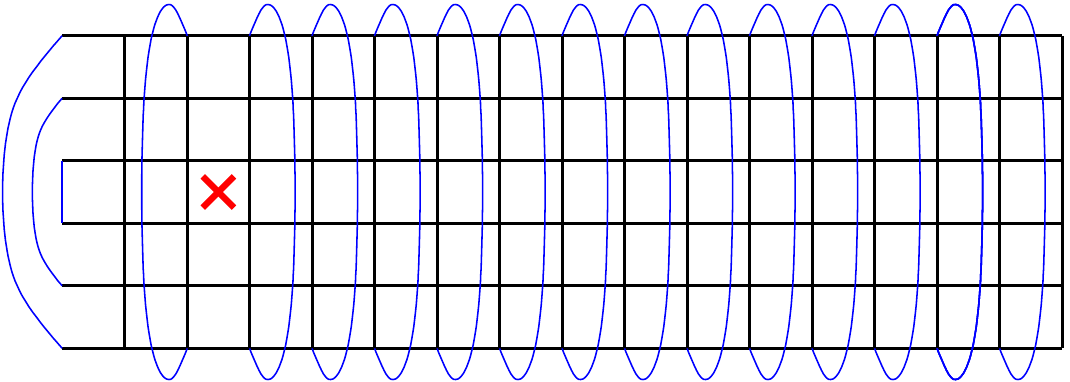} \end{center}
\caption{ (Color online) 
A point defect in 2D looks like a boundary
of an effective 1D system, if we wrap the 2D
space into a cylinder.
} 
\label{oneH} 
\end{figure}

It turns out that we may not always be able to find a  well-defined theory with
on-site symmetry in the same dimension to reproduce the low energy theory.  Let
us assume that we can always find a  well-defined gapped theory with on-site
symmetry in higher dimension to reproduce the low energy theory on a lower
dimensional defect sub-manifold, such as a boundary, a defect line, etc.  Note that
we can always deform the higher dimensional space into a lower dimensional
space so that the defect sub-manifold looks like a boundary when viewed from far
away (see Fig.  \ref{oneH}).  
So without loosing generality, we assume that  we can always find a
well-defined gapped theory with on-site symmetry in one-higher dimension to
reproduce the low energy theory on the boundary.  Therefore \frm{We can
understand anomalies through studying theories with on-site
symmetry in one-higher dimension.  }

In this paper, we will concentrate on ``pure gauge'' anomalies.  We require
that the theory is not anomalous if we break the gauge symmetry.  Within our
set up, this means that we can find a  well-defined gapped theory in same
dimension to reproduce the low energy theory, if we allow to
break the symmetry at high energies.  If we do not allow to break the 
symmetry, we still need to go to one-higher dimension. However, the
fact that the boundary theory can be well defined within the boundary (if we
break the symmetry) implies that the ground state in one-higher dimensional
theory has a trivial (intrinsic) topological
order.\cite{Wtop,Wrig,CGW1038} This way, we conclude that
\frm{We can understand ``pure'' gauge anomalies through studying SPT 
states\cite{CLW1141,CGL1172,CGL1204}
with on-site symmetry in one-higher dimension.}
A non-trivial SPT state in $(d+1)$-dimensions will correspond to a ``pure''
gauge anomaly $d$-dimensions. (For more detailed discussions, 
see section \ref{gaSPT})

With such a connection between  gauge anomalies and SPT states, we see that the
topological invariants for $(d+1)$-dimensional SPT states\cite{LG1220,W1375}
can be used to characterize $d$-dimensional gauge anomalies.  The topological
invariants for $(d+1)$-dimensional SPT states also give rise to anomaly
cancellation conditions: Given a potentially anomalous gauge theory in
$d$-dimensional space-time, we first construct a well defined
$(d+1)$-dimensional theory which produce the $d$-dimensional  gauge theory.
(This step is needed since the potentially anomalous gauge theory may not be
well defined in $d$-dimensional space-time.) If all the topological
invariants for the $(d+1)$-dimensional theory are trivial, then the
$d$-dimensional gauge theory is not anomalous.

\subsection{Gauge anomalies and gauge topological term in $(d+1)$ dimensions}

In addition to the topological invariants studied in \Ref{LG1220,W1375}, we can
also characterize gauge anomalies through the induced gauge topological term
$W^\text{gauge}_\text{top}(A_\mu)$ in the  $(d+1)$ dimensional theory, obtained
by integrating out the matter fields. The gauge topological term provide us a
powerful tool to study  gauge anomalies in one-lower dimension.

The above describes the general strategy
that we will follow in this paper.  In the following, we will first use this
line of thinking to examine several simple examples of nonABJ gauge anomalies.

\section{Simple examples of nonABJ gauge anomalies} \label{exmp}

\subsection{Bosonic $Z_2$ gauge anomaly in 1+1D}

The simplest example of nonABJ gauge anomaly is the the $Z_2$ gauge anomaly in
1+1D.  Since $\fH_\pi^3(BZ_2,\RZ)= \cH^3(Z_2,\RZ)=\Z_2$, we find that there is
only one type of non-trivial bosonic $Z_2$ gauge anomaly in 1+1D.

To see a concrete example of $Z_2$ gauge anomaly, let us first give a concrete
example of non-on-site $Z_2$ symmetry.  Gauging the non-on-site $Z_2$ symmetry
will produce the $Z_2$ gauge anomaly.

Let us consider the following spin-1/2
Ising-like model on a 1D lattice whose sites form a ring and are labeled by
$i=1,2,\cdots L$:\cite{CW1217,CLW1141,GW1248}
\begin{align}
\label{HCZX}
 H_\text{ring} &=-\sum_{i=1}^L J_{i,i+1} \si^z_i \si^z_{i+1} 
-\sum_{i=1}^L h^x_i (\si^x_i-\si^z_{i-1}\si^x_i \si^z_{i+1})
\nonumber\\
&\ \ \ 
-\sum_{i=1}^L h^y_i (\si^y_i+\si^z_{i-1}\si^y_i \si^z_{i+1})
\end{align}
where $\si^x,\si^y,\si^z$ are 2-by-2 Pauli matrices.  The model has a
non-on-site (or anomalous) $Z_2$ global symmetry generated by
\begin{align}
 U=
\prod_{i=1}^L \si^x_i
\prod_{i=1}^L \ga_{i,i+1}
\end{align}
where $\ga_{i,j}$ is a 4-by-4 matrix that acts on two spins at site-$i$ and
site-$j$ as $\ga=|{\up\up}\>\<{\up\up}| - |{\up\down}\>\<{\up\down} |+
|{\down\up}\>\<{\down\up} |+ |{\down\down}\>\<{\down\down}|$.  We say $U$ is a
non-on-site symmetry transformation since it cannot be written in the direct
product form (\ie the on-site form)\cite{CW1217,CLW1141,CGL1172,CGL1204} $
U=\otimes_i U_i$, where $U_i$ acts only on site-$i$.

Such a non-on-site (or anomalous) symmetry is not ``gaugable''. If we try to
gauge the $Z_2$ symmetry, we will get an anomalous $Z_2$ gauge theory in 1+1D.
The anomalous $Z_2$ gauge theory is not well defined and we cannot even write
down its Hamiltonian.  However, the anomalous 1+1D $Z_2$ gauge theory can be
defined as a boundary of a 2+1D $Z_2$ gauge theory.  So we can study the
physical properties of anomalous 1+1D $Z_2$ gauge theory through its
corresponding 2+1D $Z_2$ gauge theory.  We will do this in the next section for
the more general anomalous 1+1D $Z_n$ gauge theory.

In the rest of this section, we will not gauge the $Z_2$ symmetry.  We will
only study the 1+1D model with the non-on-site (\ie anomalous) $Z_2$ symmetry.
We like to understand the special properties of the 1+1D model that reflect the
anomaly (the non-on-site character) in the $Z_2$ symmetry.  

The most natural way to probe the gauge anomaly is to measure the gauge charge
induced by gauge flux.  So to probe the $Z_2$ anomaly, we like to add an unit
of $Z_2$ flux through the ring on which the 1+1D system is defined, and then
measure the induced $Z_2$ charge.  But since the $Z_2$ symmetry is non-on-site,
we do not not know how to add an unit of $Z_2$ flux through the ring.  We can
add $Z_2$ flux only if we view our anomalous 1+1D system as a boundary of a
2+1D $Z_2$ gauge theory, which will be discussed in the next section.  So here,
we will do the next best thing: we will study  the 1+1D model on a open line.
The 1+1D model on an open line can be viewed as having a strong fluctuation in
the  $Z_2$ flux through the ring.

The Hamiltonian on an open line, $H_\text{line}$, can be obtained from that on
a ring \eq{HCZX} by removing all the ``non-local terms'' that couple the site-1
and site-$L$, \ie by setting $J_{L,1}=h_1^x=h_L^x=h_1^y=h_L^y=0$.
$H_\text{line}$ still has the anomalous (\ie non-on-site) $Z_2$ symmetry: $U
H_\text{line} U^{-1}= H_\text{line}$.  However, as a symmetry transformation on
a line, $U$ contains a non-local term $\ga_{L,1}$.  After dropping the
non-local term $\ga_{L,1}$, we obtain 
\begin{align}
U_\text{line}= U \ga_{L,1}^{-1} =
\prod_{i=1}^L \si^x_i
\prod_{i=1}^{L-1} \ga_{i,i+1}.
\end{align}
We find that $U_\text{line}$ is also a symmetry of $H_\text{line}$:
$U_\text{line} H_\text{line} U_\text{line}^{-1}= H_\text{line}$.  From the
relation
\begin{align}
 U^2=U_\text{line}^4=1,\ \ \ \ U U_\text{line} U^{-1}= U_\text{line}^{-1},
\end{align}
we find that $U$ and $U_\text{line}$ generate a dihedral group $D_4=Z_4 \rtimes
Z_2$ -- a symmetry group of $H_\text{line}$.  In fact,  $H_\text{line}$ has
even higher symmetries since $\si^z_1$ and $\si^z_L$ are separately conserved.
So the full symmetry group is generated by $(\ii\si^z_1,\ii
\si^z_L,U_\text{line}, U)$ which is $[(Z_4 \times Z_4)\rtimes Z_4]\rtimes Z_2$.
Some of the group elements have the relation
\begin{align}
 (\ii \si^z_1) U = -U (\ii \si^z_1).
\end{align}
So all the representations of the group must be even dimensional.  Such a
symmetry causes a two-fold degeneracy for all the eigenvalues of
$H_\text{line}$.  From a numerical calculation, we find that the two-fold
degenerate states always carry \emph{opposite $Z_2$ quantum numbers $U=\pm 1$}.
This is a property that reflects the anomaly in the $Z_2$ symmetry.  \frm{The
two-fold degeneracy induced by the $Z_2$ non-on-site symmetry implies
that there is a Majorana zero-energy mode at each end of 1+1D system if the
system lives on an open line.}

\subsection{Bosonic $Z_n$ gauge anomalies in 1+1D}

Now let us discuss more general $Z_n$ gauge anomaly in 1+1D \emph{bosonic}
gauge theory, which is classified by $\fH_\pi^3(BZ_n,\RZ)= \cH^3(Z_n,\RZ)=\Z_n$.
So there are $n-1$ non-trivial  $Z_n$ gauge anomalies.  To construct the
examples of those $Z_n$ gauge anomalies, we will present two approaches here.  

In the first approach, we start with a bosonic $Z_n$ SPT state in 2+1D.  We can
realize the $Z_n$ SPT state through a 2+1D bosonic $U(1)$ SPT state, which is
described by the following $U(1)\times U(1)$ Chern-Simons
theory:\cite{LV1219,SL1204}
\begin{align}
\label{CS}
\cL=
\frac{1}{4\pi} K_{IJ} a_{I\mu}\prt_\nu a_{J\la}\eps^{\mu\nu\la}
+
\frac{1}{2\pi} q_{I} A_{\mu}\prt_\nu a_{I\la}\eps^{\mu\nu\la}
+\cdots
\end{align}
where the non-fluctuating probe field $A_\mu$ couples to the current of the
global $U(1)$ symmetry.  Here the $K$-matrix and the charge vector $\v q$ are
given by\cite{BW9045,R9002,WZ9290}
\begin{align}
\label{Kq}
K=
\begin{pmatrix}
0 & 1 \\
1 & 0 \\
\end{pmatrix}
,\ \ \ \ 
\v q=
\begin{pmatrix}
1  \\
k  \\
\end{pmatrix}
, \ \ \ k \in \Z.
\end{align}
The even diagonal elements of the $K$-matrix are required by the bosonic nature
of the theory.  The Hall conductance for the $U(1)$ charge coupled to $A_\mu$
is given by
\begin{align}
 \si_{xy}= (2\pi)^{-1} \v q^T K^{-1} \v q = \frac{2k}{2\pi}.
\end{align}

The above 2+1D $U(1)$ SPT state is characterized by an integer $k \in
\cH^3[U(1),\RZ]$.  We also know that an  2+1D $Z_n$ SPT state is characterized
by a mod-$n$ integer $m \in \cH^3(Z_n,\RZ)$.  If we view the 2+1D $U(1)$ SPT
state labeled by $k$ as a 2+1D $Z_n$ SPT state, then what is the $m$ label for
such a 2+1D $Z_n$ SPT state?

The  mod-$n$ integer $m$ can be measured through a topological invariant
constructed by creating $n$ identical $Z_n$ monodromy defects:\cite{W1375}
\emph{$2m$ is the total $Z_n$ charge of  $n$ identical $Z_n$ monodromy
defects}. On the other hand, a $Z_n$ monodromy defect corresponds to $2\pi/n$
flux in the $U(1)$ gauge field $A_\mu$.  From the $2k$ quantized Hall
conductance, $n$ identical $2\pi/n$-flux of $A_\mu$ will induce $2k$ $U(1)$
charge, which is also the $Z_n$ charge.  So the above bosonic $U(1)$ SPT state
correspond to a $m=k$ mod $n$ bosonic $Z_n$ SPT state.\cite{W1375}  

\begin{figure}[tb] 
\begin{center} 
\includegraphics[scale=0.3]{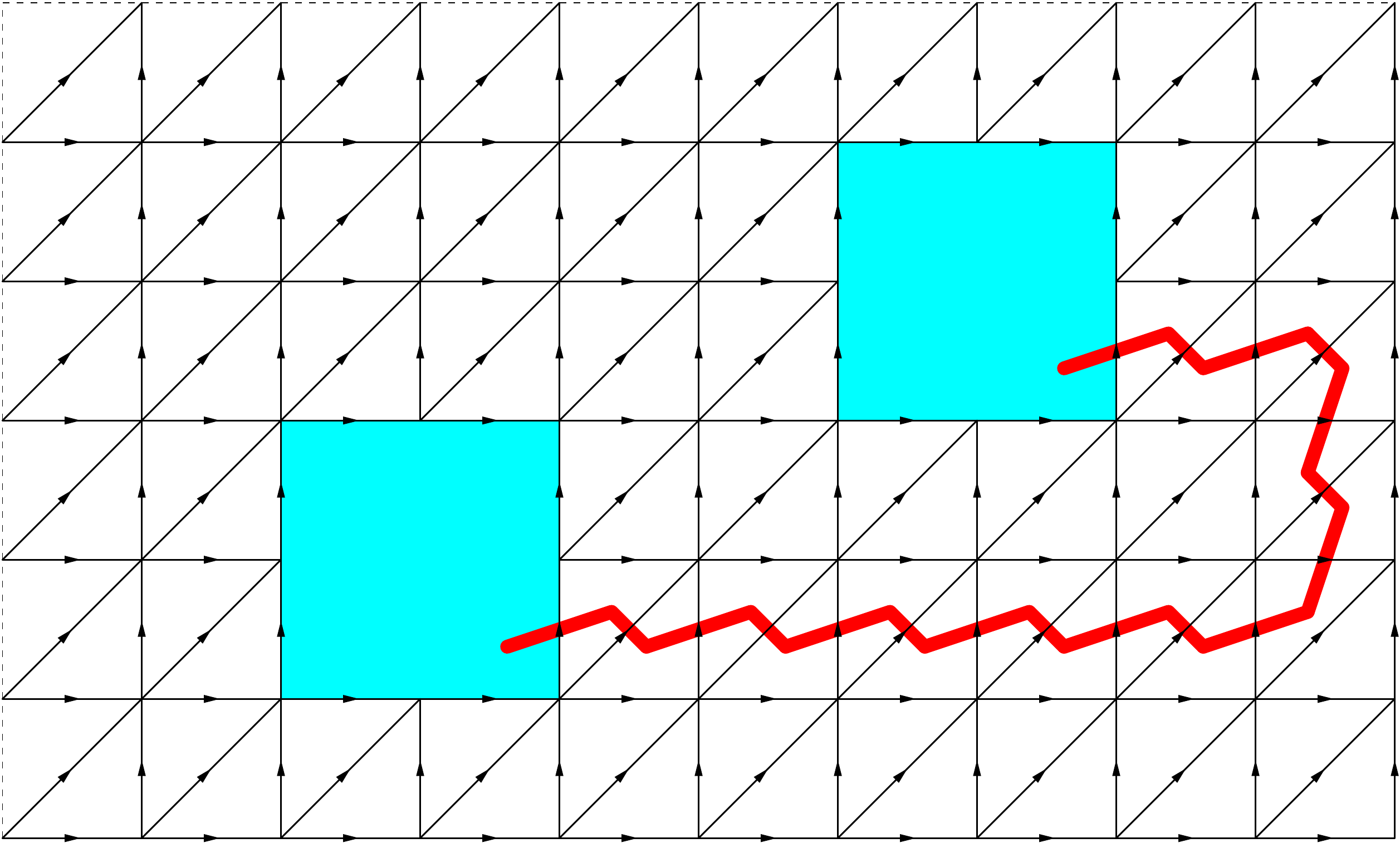} \end{center}
\caption{ (Color online) 
A $Z_2$ gauge configuration with two \emph{identical} holes on a torus that
contains a unit of $Z_2$ flux in each hole.  The $Z_2$ link variables are equal
to $-1$ on the crossed links and $1$ on other links. If the 1+1D bosonic $Z_2$
gauge theory on the edge of one hole is  anomalous, then such a $Z_2$ gauge
configuration induces half unit of total $Z_2$ charge on the edge (representing
a $Z_2$ gauge anomaly).  Braiding those holes around each other reveals the
fractional statistics of the holes.  The edge states for one hole are
degenerate with $\pm 1/2$ $Z_2$ charge if there is a time-reversal symmetry. 
} 
\label{z2gaugeH} 
\end{figure}

The low energy effective edge theory for the 2+1D system \eq{CS} has an
non-on-site $Z_n$ symmetry if $k \neq 0$ mod n.  (In fact, the low energy edge
effective theory has an non-on-site $U(1)$ symmetry.) If we gauge such a
non-on-site $Z_n$ symmetry on the edge, we will get an anomalous $Z_n$ gauge
theory in 1+1D, which is not well defined.  (In other words, we cannot gauge
non-on-site $Z_n$ symmetry within 1+1D.) 

However, we can define  an anomalous $Z_n$ gauge theory in 1+1D as the edge
theory of a 2+1D $Z_n$ gauge theory.
Such a 2+1D $Z_n$ gauge theory
can be obtained from
\eqn{CS} by treating $A_\mu$ as a dynamical $U(1)$ gauge field and
introduce a charge $n$-Higgs field to break the $U(1)$ down to $Z_n$:
\begin{align}
\label{CSZn}
\cL &=
\frac{1}{4\pi} K_{IJ} a_{I\mu}\prt_\nu a_{J\la}\eps^{\mu\nu\la}
+
\frac{1}{2\pi} q_{I} A_{\mu}\prt_\nu a_{I\la}\eps^{\mu\nu\la}
\nonumber\\
&\ \ \ \
+|(\prt_\mu +\ii n A_\mu)\phi|^2 + a |\phi|^2 - b |\phi|^4
\end{align}
The edge theory of the above Ginzberg-Landau-Chern-Simons theory contain
gapless edge excitations with central charge $c=1$ right-movers and central
charge $\bar c=1$ left-movers. Such a 1+1D edge theory is an example of
anomalous 1+1D $Z_n$ gauge theory that we are looking for.  The anomaly is
characterized by a mod-$n$ integer $m=2k$.  A unit of $Z_n$ flux (a $2\pi/n$
flux) through the hole (see Fig.  \ref{z2gaugeH}) will induce a $2m/n$ $Z_n$
charge on the edge.  Such a property directly reflects a $Z_n$ gauge anomaly.


To summarize, in the first approach, we start with a $Z_n$ SPT state in 2+1D to
produce a 1+1D edge theory with a non-on-site $Z_n$ symmetry.  We then gauge
the non-on-site $Z_n$ symmetry to obtain an anomalous $Z_n$ gauge theory in
1+1D.

In the second approach, we use the the Levin-Gu duality
relation\cite{LG1220,HW1267,HWW1295} between the $Z_n$ SPT states and the
(twisted) $Z_n$ gauge theory in 2+1D.  We obtain the anomalous 1+1D bosonic
$Z_n$ gauge theory directly as the edge theory of the (twisted) $Z_n$ gauge
theory in 2+1D.  The (twisted) $Z_n$ gauge theory can be described by the
following 2+1D $U(1)\times U(1)$ Chern-Simons
theory:\cite{KLW0834,LS0903,HW1227}
\begin{align}
\label{CSZnCS}
\cL=
\frac{1}{4\pi} \t K_{IJ} a_{I\mu}\prt_\nu a_{J\la}\eps^{\mu\nu\la}
+\cdots
\end{align}
where the $\t K$-matrix is given by
\begin{align}
\label{KKinv}
\t K=
\begin{pmatrix}
-2m & n \\
n & 0 \\
\end{pmatrix}
,\ \ \ \
\t K^{-1}=
\begin{pmatrix}
0 & 1/n \\
1/n & 2m/n^2 \\
\end{pmatrix} .
\end{align}

When $m=0$, the above 2+1D theory is a standard $Z_n$ gauge theory, and its low
energy edge theory is a standard $Z_n$ gauge theory in 1+1D with no anomaly.
Such an 1+1D   $Z_n$ gauge theory can defined within 1+1D without going through
a 2+1D theory.  When $m \neq 0$, the $m$ term corresponds to a quantized
topological term in $Z_n$ gauge theory discussed in \Ref{HW1267}.  Such a
quantized topological term is classified by a mod-$n$ integer $m \in
\cH^3(Z_n,\RZ)$ 


To see the relation between the $U(1)\times U(1)$ Chern-Simons theory
\eqn{CSZnCS} and the $Z_n$ gauge theory in 2+1D,\cite{KLW0834,LS0903} we note
that a unit $a_{1\mu}$-charge correspond to a unit $Z_n$ charge.  A unit $Z_n$
charge always carries a Bose statistics.  So the $Z_n$ gauge theory is a
bosonic  $Z_n$ gauge theory.  On the other hand, a unit of $Z_n$ flux is
described by a particle with $l^v_I$ $a_{I\mu}$-charge. We find that $l^v_2=1$
(so that moving a unit $Z_n$ charge around a unit $Z_n$ flux will induce
$2\pi/n$ phase).  $l^v_1$ can be any integer and the  $\v l^v=(l^v_1,1)$
$a_{I\mu}$-charge is not a pure  $Z_n$ flux (\ie may carry some $Z_n$ charges).

When the 2+1D system \eq{CSZnCS} has holes (see Fig.  \ref{z2gaugeH}), the
theory lives on the edge of hole is an 1+1D anomalous $Z_n$ gauge theory.  If
we add $Z_n$ flux to the hole, we may view the hole as a particle with $\v
l=(0,1)$ $a_{I\mu}$-charge. Such a  particle carries  a fractional $2m/n$ $Z_n$
charge as discussed above.  We conclude that, when $m\neq 0$, \frm{a unit of
$Z_n$ flux through a ring, on which a 1+1D anomalous  $Z_n$ gauge theory lives,
always induces a fractional $Z_n$ charge $2m/n$, which is a consequence of $Z_n$
gauge anomaly of the 1+1D system.}


From the second description of the anomalous 1+1D bosonic $Z_n$ gauge
theory, we see that if we view the holes with a unit of $Z_n$ flux as particles
(see Fig.  \ref{z2gaugeH}), then such particles will carry a unit of
$a_{2\mu}$-charge.  If we braid the  holes with a unit of $Z_n$ flux around
each others (see Fig.  \ref{z2gaugeH}), those holes will carry a fractional
statistics $\th=\frac{2m}{n^2}\pi$ (the fractional statistics of unit
$a_{2\mu}$-charges).  \frm{One can use fractional (or non-Abelian) statistics of
the holes with flux to detect 1+1D gauge
anomaly.\cite{LG1220}}

The gapless edge excitations of the 2+1D theory \eq{CSZnCS}
is described by the following 1+1D effective theory
\begin{align}
\label{L1D}
 \cL_{1+1D} & =
\frac{1}{4\pi} [
\t K_{IJ} \prt_t \phi_I \prt_x \phi_J-
V_{IJ} \prt_x \phi_I \prt_x \phi_J ] 
\\
&\ \ \
+ \sum_l \sum_{J=1,2} [
c_{J,l} \ee^{\ii l \t K_{JI}\phi_I}
h.c.
]
,
\nonumber 
\end{align}
where the field $\phi_I (x,t)$ is a map from the 1+1D space-time to a circle
$2\pi \RZ$, and $V$ is a positive definite real 2-by-2 matrix.  

A $Z_n$ flux (not the pure $Z_n$ flux which is not allowed) is described by an
unit $a_{2\mu}$-charge.  From the equation of the motion, we find that, in the
bulk, a $Z_n$ flux correspond to a bound state of $1/n$ $a_{1\mu}$-flux and
$2m/n^2$ $a_{2\mu}$-flux.  Thus  a unit of $Z_n$ flux through the hole is
described by the following boundary condition\cite{W9211,W9505}
\begin{align}
 \phi_1(x) &= \phi_1(x+L)+2\pi/n, 
\nonumber\\
 \phi_2(x) &= \phi_2(x+L)+2\pi(2m/n^2) .
\end{align}
We see that the $Z_n$ symmetry of the 1+1D theory is generated by
\begin{align}
\label{Zntrns}
 \phi_1 \to \phi_1  +2\pi/n ,\ \ \
 \phi_2 \to \phi_2  +4\pi m/n^2 .
\end{align}
Such a $Z_n$ symmetry is anomalous (or non-on-site)
if $m\neq 0$ mod $n$ in \eqn{KKinv}.  When
$n=2$ and $m=1$, \eqn{L1D} is the low energy effective theory of
$H_\text{ring}$ in \eqn{HCZX}.  

If we gauge the $Z_n$ symmetry, we will get an anomalous $1+1D$ $Z_n$ gauge
theory, which has no 1+1D non-perturbative definition.  This way, we obtain an
example of bosonic anomalous $Z_n$ gauge theory in 1+1D, \eqn{L1D} and
\eqn{Zntrns}.

\subsection{Bosonic $Z_2\times Z_2\times Z_2$ gauge anomalies in 1+1D}

The bosonic $Z_2\times Z_2\times Z_2$ gauge anomalies in 1+1D are classified
by $\fH_\pi^3(B(Z_2\times Z_2\times Z_2),\RZ)= \cH^3[Z_2\times Z_2\times Z_2,\RZ]
=\Z_2^7$.  So there are 127 different types of $Z_2\times Z_2\times Z_2$ gauge
anomalies in 1+1D.  Those 127 gauge anomalies in 1+1D can be constructed by
starting with a 2+1D  $Z_2\times Z_2\times Z_2$ gauge theory.  We then add the
quantized topological terms\cite{HW1267} to twist the $Z_2\times Z_2\times Z_2$
gauge theory.  The quantized topological terms are also classified by
$\cH^3[Z_2\times Z_2\times Z_2,\RZ]$.  The low energy edge theories of those
twisted  $Z_2\times Z_2\times Z_2$ gauge theories realize the 127 types of
bosonic $Z_2\times Z_2\times Z_2$ gauge anomalies in 1+1D.  The edge theories
always have degenerate ground states or gapless excitations, even after we
freeze the $Z_n$ gauge fluctuations (\ie treat the $Z_n$ gauge field as a
non-dynamical probe field).

As discussed in \Ref{JWW}, 64 twisted  $Z_2\times Z_2\times Z_2$ gauge
theories can be described by $U^6(1)$ Chern-Simons theories \eq{CSZnCS} with
\begin{align}
\t K=  \begin{pmatrix}
  -2m_1 & 2 &-m_{12}& 0 &-m_{13} & 0 \\
    2   & 0 & 0      & 0 & 0       & 0 \\
-m_{12}& 0 &-2m_2   & 2 &-m_{23} & 0 \\
    0   & 0 & 2      & 0 & 0       &  0\\
-m_{13}& 0 &-m_{23}& 0 &-2m_3    & 2 \\
    0   & 0 & 0      & 0 &   2     & 0 \\
\end{pmatrix} ,
\end{align}
where $m_i,m_{ij}=0,1$. The $m_i$ terms and the $m_{ij}$ terms are the
quantized topological terms, which twist the $Z_2\times Z_2\times Z_2$ gauge
theory.  The other 64 twisted  $Z_2\times Z_2\times Z_2$ gauge theories are
non-Abelian gauge theories\cite{P9797,JWW}  with gauge groups $D_4$, $Q_8$,
\etc (see also \Ref{MR1235}).  So some of the anomalous  $Z_2\times Z_2\times
Z_2$ gauge theories in 1+1D has to be defined via non-Abelian gauge theories in
2+1D.  (For details, see \Ref{P9797,JWW}.) In this case, the holes that carry
the gauge flux have non-Abelian statistics (see Fig. \ref{z2gaugeH}).

\subsection{Fermionic $Z_2\times Z_2$ gauge anomalies in 1+1D}

A fermionic $Z_2\times Z_2$ anomalous (\ie non-on-site) symmetry in 1+1D can be
realized on the edge of a 2+1D fermionic $Z_2\times Z_2$ SPT states.  Those
fermionic SPT states were discussed in detail in \Ref{GW1248,RZ1232,W1375}.  We
found that there are 16 different fermionic $Z_2\times Z_2$ SPT states in 2+1D
(including the trivial one) which form a $\Z_8$ group where the
group operation is the stacking of the 2+1D states.


One type of the fermionic $Z_2\times Z_2$ anomalous  symmetry in 1+1D is
realized by the following free Majorana field theory
\begin{align}
\label{MajF}
 \cL_{1+1D}=
\ii \la_R (\prt_t -\prt_x) \la_R +
\ii \la_L (\prt_t +\prt_x) \la_L .
\end{align}
The $Z_2\times Z_2$ symmetry is generated by
the following two generators
\begin{align}
  (\la_R, \la_L) \to (-\la_R, \la_L) ,\ \ \ 
  (\la_R, \la_L) \to (\la_R, -\la_L) ,
\end{align}
\ie $\la_R$ carries the first $Z_2$ charge and $\la_L$ the second $Z_2$ charge.
The the above fermionic anomalous  symmetry is the generator of $\Z_8$
types of fermionic $Z_2\times Z_2$ anomalous symmetries.

Due to the anomaly in the  $Z_2\times Z_2$ symmetry, the above 1+1D field
theory can only be realized as a boundary of a 2+1D lattice model if we require
the $Z_2\times Z_2$ symmetry to be an on-site symmetry.  (However, it may be
possible to realize the  1+1D field theory by a 1+1D lattice model if we do not
require the $Z_2\times Z_2$ symmetry to be an on-site symmetry.) One example of
2+1D realization is the stacking of a $p+\ii p$ and a $p-\ii p$ superconductor
(denoted as $p+\ii p$/$p-\ii p$ state).\cite{RG0067,RZ1232,W1375}  

Since the  $Z_2\times Z_2$ symmetry is anomalous in the above 1+1D field
theory, if we gauge the $Z_2\times Z_2$ symmetry, the resulting 1+1D fermionic
$Z_2\times Z_2$ gauge theory will be anomalous which can not have a
non-perturbative definition as a 1+1D model.  However, the  1+1D fermionic
$Z_2\times Z_2$ gauge theory can have a non-perturbative definition as the
boundary of a 2+1D model.  One such model is the stacking of a bosonic $\nu=1$
Pfaffian quantum Hall state\cite{MR9162} and a bosonic $\nu=-1$ Pfaffian
quantum Hall state (denoted as Pfaff/$\overline{\text{Pfaff}}$ state).  Note
that the  bosonic $\nu=1$ Pfaffian quantum Hall state have edge states which
include a $c=1/2$ Majorana mode and a $c=1$ density mode.\cite{W9355} However,
since we do not require boson number conservation, the density mode of the
$\nu=1$ Pfaffian state and the density mode of the $\nu=-1$ Pfaffian state can
gap out each other, and be dropped.

Again, it is interesting to see that a non-perturbative definition of an
anomalous 1+1D fermionic  $Z_2\times Z_2$ Abelian gauge theory requires an
non-Abelian state\cite{W9102,MR9162} in 2+1D.


\subsection{Bosonic $U(1)$ gauge anomalies in 2+1D}
\label{U1B3d}

The bosonic $U(1)$ gauge anomalies in 2+1D are described by
$\fH_\pi^4[BU(1),\RZ]$ which contains $H^4[BU(1),\RZ]=\RZ$ as subgroup.  So
what are those $U(1)$ gauge anomalies labeled by a real number $\ka/2 \in
\RZ=[0,1)$?  

First, let us give a more general definition of anomalies (which include gauge
anomalies as special cases): We start with a description of a set of low energy
properties, and then ask if the set of low energy properties can be realized by
a well-defined quantum theory in the same dimensions?  If not, we say the
theory is anomalous.

So to describe the 2+1D $U(1)$ gauge anomaly, we need to first describe a set
of low energy properties.  The $U(1)$ gauge anomaly is defined by the following
low energy properties:\\ 
(1) there are no gapless excitations and no ground state degeneracy.\\ 
(2) the $U(1)$ gauge theory has a fractional Hall conductance 
$\si_{xy}=\ka/2\pi$.\\ 
The above  low energy properties implies
that, after integrating out the matter field, the 2+1D  theory produces the
following gauge topological term
\begin{align}
 \cL_{2+1D}=
\frac{\ka}{4\pi} A_\mu\prt_\nu A_\la \eps^{\mu\nu\la}+ \cdots .
\end{align}
When  $\ka \in [0,2)$, the above  two  low energy properties cannot be realized
by  a well-defined local bosonic quantum theory in 2+1D.  In this case, the
theory has a $U(1)$ gauge anomaly.

To see the above two properties cannot be realized by a well-defined 2+1D
bosonic theory (\ie represent a $U(1)$ gauge anomaly), we first note that the
requirement that there is no degenerate ground
states implies that there are no excitations with fractional charges and
fractional statistics (since the state has no intrinsic topological
order\cite{Wtop,Wrig}).  Second, the above $U(1)$ Chern-Simons theory with a
\emph{fractional coefficient} has a special property that a unit of $U(1)$ flux
($2\pi$ flux) induces a $U(1)$ charge $\ka$ (since the Hall conductance is
$\frac{\ka}{2\pi}$).  The flux-charge bound state has a statistics
$\th=\ka \pi$.  Since  a unit of $U(1)$ flux only induce an allowed
excitation, so for any  well-defined 2+1D model with no ground state
degeneracy, the induced charge must be integer, and the induced statistics must
be bosonic (for a bosonic theory):
\begin{align}
 \ka = \text{ integer}, \ \ \ \
 \ka = \text{ even integer} .
\end{align}
We see that, for $\ka \in [0,2)$, the above  $U(1)$ Chern-Simons theory (with
no ground state degeneracy) cannot appear as the low energy effective theory of
any well-defined 2+1D model.  Thus, it is anomalous.

But when $\ka=$ even integer, the above 2+1D model with even-integer quantized
Hall conductance can be realized through a well-defined 2+1D bosonic model with
trivial topological order,\cite{LV1219,LW1224,CW1217,SL1204} and thus not
anomalous.\cite{VS1258} This is why only $\ka \in [0,2)$ represents the $U(1)$
anomalies in 2+1D.

However, the above anomalous 2+1D theory (with no ground state degeneracy) can
be realized as the boundary theory of a 3+1D bosonic insulator that does not
have the time-reversal and parity symmetry.  The 3+1D bosonic insulator
contains a topological term
\begin{align}
\cL_{3+1D}= \frac{2\pi \ka}{2! (2\pi)^2} \prt_\mu A_\nu \prt_\la A_\ga
\eps^{\mu\nu\la\ga}
\end{align}
that is allowed by symmetry.  A unit of magnetic flux through the boundary will
induce a fractional $U(1)$ charge $\ka$ on the boundary.  Thus the 3+1D bosonic
insulator can reproduces the above two mentioned low energy properties.

The above result can be generalized to study $U^k(1)$ gauge anomaly in 2+1D.
If after integrating out the matter fields, we obtain
the following gauge topological term
\begin{align}
 \cL_{2+1D}=
\frac{\ka_{IJ}}{4\pi} A^I_\mu\prt_\nu A^J_\la \eps^{\mu\nu\la}+ \cdots ,
\end{align}
then the theory is anomalous if $\ka_{IJ}$ is not an integer
symmetric matrix with even diagonal elements.
The  anomalous $U^k(1)$ gauge theory can be viewed as the boundary 
of a 3+1D $U^k(1)$ gauge theory with topological term
\begin{align}
\cL_{3+1D}= \frac{2\pi \ka_{IJ}}{2! (2\pi)^2} \prt_\mu A^I_\nu \prt_\la A^J_\ga
\eps^{\mu\nu\la\ga} .
\end{align}
Two topological terms described by $\ka_{IJ}$ and $\ka'_{IJ}$ are regarded as
equivalent if
\begin{align}
 \ka'_{IJ} - \ka_{IJ} = K^\text{even}_{IJ}
\end{align}
where $K^\text{even}$ is an integer symmetric matrix with even diagonal
elements.\cite{MKF1335}

\subsection{Fermionic $U(1)$ gauge anomalies in 2+1D}

In this section, we consider 2+1D fermion systems with a $U(1)$ symmetry where
\emph{the fermion parity symmetry is part of $U(1)$ symmetry}.  As a result,
all fermions carry odd $U(1)$ charges.  

The gauge anomalies in such fermionic $U(1)$ gauge theory are described by
$\fH_\pi^4[BU(1),\RZ]$ which includes $H^4[BU(1),\RZ]=\RZ$.  Thus, the fermionic
$U(1)$ gauge anomaly can be labeled by $\ka \in \RZ = [0,1)$.

The  $U(1)$ gauge anomaly correspond to the following low energy properties:
The 2+1D anomalous fermionic $U(1)$ gauge theory has\\
(1) a gapped non-degenerate ground state and \\
(2) a fractional Hall conductance $\si_{xy}=\ka/2\pi$.\\
After integrating out the matter field, the 2+1D  theory produces the
following gauge topological term
\begin{align}
 \cL_{2+1D}=
\frac{\ka}{4\pi} A_\mu\prt_\nu A_\la \eps^{\mu\nu\la}+ \cdots .
\end{align}
When $\ka \in [0,1)$ the above 2+1D theory is anomalous.

When $\ka=$ integer, the above 2+1D model with integer quantized Hall
conductance can be realized through a well-defined 2+1D fermionic model -- an
integer quantized Hall state which has no ground state degeneracy. So the 2+1D
theory with integer $\ka$ is not anomalous. This is why only $\ka \in [0,1)$
represents the fermionic $U(1)$ anomalies in 2+1D.
 
Similarly, the above result can also be generalized to study fermionic $U^k(1)$
gauge anomaly in 2+1D.  If after integrating out the matter fields, we obtain
the following gauge topological term
\begin{align}
 \cL_{2+1D}=
\frac{\ka_{IJ}}{4\pi} A^I_\mu\prt_\nu A^J_\la \eps^{\mu\nu\la}+ \cdots ,
\end{align}
then the theory is anomalous if and only if $\ka_{IJ}$ is not an integer
symmetric matrix.
The  anomalous fermionic $U^k(1)$ gauge theory can be viewed as the boundary 
of a 3+1D $U^k(1)$ gauge theory with topological term
\begin{align}
\cL_{3+1D}= \frac{2\pi \ka_{IJ}}{2! (2\pi)^2} \prt_\mu A^I_\nu \prt_\la A^J_\ga
\eps^{\mu\nu\la\ga} .
\end{align}
Two topological terms described by $\ka_{IJ}$ and $\ka'_{IJ}$ are regarded as
equivalent if
\begin{align}
 \ka'_{IJ} - \ka_{IJ} = K_{IJ}
\end{align}
where $K$ is an integer symmetric matrix.  It is interesting to see that the
periodicy of $ \ka_{IJ}$ is an even integer matrix for bosonic systems while
the periodicy is an integer matrix for fermionic
systems.\cite{MKF1335}

\subsection{$U(1)\times [U(1)\rtimes Z_2]$ gauge anomalies in 2+1D}

After understanding the $U(1)$ gauge anomalies in 2+1D for bosonic and
fermionic systems, we are ready to discuss a more interesting example --
$U(1)\times [U(1)\rtimes Z_2]$ gauge anomalies in 2+1D.

\subsubsection{Cohomology description}

The 2+1D $U(1)\times [U(1)\rtimes Z_2]$ gauge anomalies are described by
$\fH_\pi^4[B(U(1)\times [U(1) \rtimes Z_2]),\RZ]$ which contains $H^4[B(U(1)\times
[U(1)\rtimes Z_2]),\RZ]$ as a subgroup.  Using K\"unneth formula (see 
\eqn{kunnU}), we can compute $H^d[B(U(1)\times [U(1)\rtimes Z_2]),\Z]$ from $H^d[
B[U(1)\rtimes Z_2],\Z]$ and $H^d[BU(1),\Z]$:
\begin{widetext}
\begin{align}
\begin{matrix}
d:                          &0 ,&1,& 2  ,& 3  ,& 4          ,& 5  ,& 6 \\
H^d[B[U(1)\rtimes Z_2],\Z]:&\Z,&0,&\Z_2,&\Z_2,&\Z\oplus\Z_2,&\Z_2,&\Z_2^{\oplus 2} \\ 
H^d[ BU(1)            ,\Z]:&\Z,&0,&\Z  ,& 0  ,&\Z          ,&0   ,&\Z \\ 
H^d[B(U(1)\times [U(1)\rtimes Z_2]),\Z]:
&\Z ,& 0 ,& \Z\oplus\Z_2 ,& \Z_2 ,& \Z^{\oplus 2}\oplus\Z_2^{\oplus 2} ,& \Z_2^{\oplus 2} ,& \Z^{\oplus 2}\oplus\Z_2^{\oplus 4}  \\ 
\end{matrix}
\end{align}
where $\Z_n^{\oplus 2}\equiv \Z_n\oplus \Z_n$.  Then using the universal
coefficient theorem  (see appendix \ref{KF}), we find
\begin{align}
\begin{matrix}
d:                     &0 ,&1,& 2  ,& 3  ,& 4          ,& 5   \\
H^d[B(U(1)\times [U(1)\rtimes Z_2]),\RZ]:
&\RZ ,& Z_2 ,& \RZ\oplus\Z_2 ,& \Z_2^{\oplus 2} ,& (\RZ)^{\oplus 2}\oplus\Z_2^{\oplus 2} ,& \Z_2^{\oplus 4}   \\ 
\end{matrix}
\end{align}
\end{widetext}
We see that some of the $U(1)\times [U(1)\rtimes Z_2]$ gauge anomalies in 2+1D
can be described by $(\RZ)^{\oplus 2}\oplus\Z_2^{\oplus 2}
\subset \fH_\pi^4[B(U(1)\times [U(1) \rtimes Z_2]),\RZ]$.  

\subsubsection{Continuous gauge anomalies}

The gauge anomalies described by $(\RZ)^{\oplus 2}$ can be labeled by two real
numbers $(\ka_1,\ka_2)\in (\RZ)^{\oplus 2}$ (for fermions) or $(\ka_1,\ka_2)
\in (\R/2\Z)^{\oplus 2}$ (for bosons).  An example of such a  gauge anomaly can
be obtained through a  $U(1)\times [U(1)\rtimes Z_2]$ gauge theory coupled to
matter fields.  If integrating out the matter field produces the following
gauge topological term in 3+1D:
\begin{align}
\cL_{3+1D} &= 
\frac{2\pi \ka_1}{2! (2\pi)^2} \prt_\mu A_{1\nu} \prt_\la A_{1\ga}\eps^{\mu\nu\la\ga} 
\nonumber\\
&
+ \frac{2\pi \ka_2}{2! (2\pi)^2} \prt_\mu A_{2\nu} \prt_\la A_{2\ga} \eps^{\mu\nu\la\ga}
\end{align}
then the 3+1D gauge theory describes the desired  gauge anomaly.  Here
$A_{1\mu}$ is for the first $U(1)$ and $A_{2\mu}$ the second $U(1)$, and
$A_{2\mu}$ changes sign under the $Z_2$ gauge transformation.

\subsubsection{First discrete gauge anomaly}
\label{U1U1Z2}

If integrating out the matter field produces the following gauge topological
term in 3+1D:\cite{HW1267}
\begin{align}
\label{fUUZ}
\cL_{3+1D}= 
\frac{\pi}{(2\pi)^2} \prt_\mu A_{1\nu} \prt_\la A_{2\ga} \eps^{\mu\nu\la\ga}
\end{align}
then the 3+1D gauge theory describes a discrete  $U(1)\times[ U(1) \rtimes
Z_2]$ gauge anomaly (which belongs to $\Z_2^{\oplus 2}$).  The boundary 2+1D
theory of the 3+1D system will be a  $U(1)\times [U(1)\rtimes Z_2]$ gauge
theory with the discrete  $U(1)\times [U(1)\rtimes Z_2]$ gauge anomaly.  Such
an anomalous 2+1D theory must be gapless or have degenerate ground states, if
we freeze the gauge fluctuations without breaking the $U(1)\times( U(1) \rtimes
Z_2)$ symmetry.  We suspect that, in our particular case,  2+1D boundary
theory is actually gapless.

This is because if the $Z_2$ gauge symmetry is broken on the 2+1D boundary, we
will have the following effective 2+1D boundary theory:
\begin{align}
\cL=
\frac{\phi/|\phi|}{4\pi} \t K_{IJ} A_{I\mu}\prt_\nu A_{J\la}\eps^{\mu\nu\la}
+\cdots
\end{align}
where $\phi$ is the Higgs field that breaks the $Z_2$ gauge symmetry, and the
$\t K$-matrix is given by
\begin{align}
\label{KKinv}
\t K=
\begin{pmatrix}
0 & 1/2 \\
1/2 & 0 \\
\end{pmatrix}
.
\end{align}
The above theory has a fractional mutual Hall conductance
\begin{align}
 \si^{IJ}_{xy}=\frac{\phi}{|\phi|}\frac{\t K_{IJ}}{2\pi} , \ \ 
 \si^{11}_{xy} = \si^{22}_{xy} = 0, \ \ 
 \si^{12}_{xy} = \si^{21}_{xy} = \frac{1/2}{2\pi}\frac{\phi}{|\phi|}
.
\end{align}
Such a theory can be realized by a double-layer bosonic fractional quantum Hall
state described by $K$-matrix $ K=\frac{\phi}{|\phi|}
\begin{pmatrix}
0 & 2 \\
2 & 0 \\
\end{pmatrix}$
where the bosons in the two layers carry unit charges of the two $U(1)$'s
separately.

If the $Z_2$ gauge symmetry does not break,  $\phi$ will fluctuate with
equal probability to be $\phi=\pm|\phi|$.  Due to the separate
conservation of the two $U(1)$ charges, the domain wall between $\phi=+|\phi|$
and $\phi=-|\phi|$ will support gapless edge excitations.\cite{W9125,W9211}
Because there are long  domain walls in the disordered phase of $\phi$, this
suggests that the theory is gapless if the $U(1)\times [U(1)\rtimes Z_2]$
symmetry is not broken.

To further understand the physical property of such a discrete gauge
anomaly, let us assume that the 3+1D space-time has a topology $M_2\times
M_2'$.  We also assume that the $A_{1\mu}$ gauge field has $2\pi$ flux on
$M_2'$.  In the large $M_2$ limit, the Lagrangian \eq{fUUZ} reduces to an
effective Lagrangian on  $M_2$ which has a form
\begin{align}
\cL_{M_2}= 
\frac{\pi}{2\pi} \prt_\mu A_{2\nu} \eps^{\mu\nu}
.
\end{align}
We note that the  $A_{1\mu}$ gauge configuration preserve the  $U(1)\times( U(1)
\rtimes Z_2)$ symmetry. The above Lagrangian is the effective  Lagrangian of the
$U(1)\times [U(1)\rtimes Z_2]$ symmetric theory on $M_2$ probed by the
$A_{2\mu}$ gauge field.\cite{W1375} Such an effective  Lagrangian implies that
the $U(1)\times [U(1)\rtimes Z_2]$ symmetric theory on $M_2$ describe a
non-trivial $U(1)\times [U(1)\rtimes Z_2]$ SPT state labeled by the non-trivial
element in $\cH^2[U(1)\times (U(1) \rtimes Z_2),\RZ]=\Z_2$.\cite{HW1227}

The non-trivial  1+1D $U(1)\times [U(1)\rtimes Z_2]$ SPT state on $M_2$ has the
following property: Let $M_2=R_t\times I$, where $R_t$ is the time and $I$ is a
spatial line segment. Then the excitations at the end of the line are
degenerate, and the degenerate end-states form a projective representation of
$U(1)\times[U(1)\rtimes Z_2]$.\cite{CGW1107,SPC1139,CGW1128,PBT1039}

The above result has another interpretation.  Let the 3+1D space-time has a
topology $R_t\times I\times M_2'$.  Such a space-time has two boundaries.  Each
boundary has a topology $R_t\times M_2'$, and the theory on the boundary is a
$U(1)\times [U(1)\rtimes Z_2]$ gauge theory with the first discrete $U(1)\times
[U(1)\rtimes Z_2]$ gauge anomaly.  If we freeze the  $U(1)\times [U(1)\rtimes
Z_2]$ gauge fields without break the $U(1)\times [U(1)\rtimes Z_2]$ symmetry,
then \emph{all the low energy excitations} on $M_2'$ at one boundary form a
linear representation of $U(1)\times [U(1)\rtimes Z_2]$, if the $A_{1\mu}$
gauge field is zero on $M_2'$.  However, \emph{all the low energy excitations}
on $M_2'$ at one boundary will form a projective representation of $U(1)\times
[U(1)\rtimes Z_2]$, if the $A_{1\mu}$ gauge field has $2\pi$ flux on $M_2'$.
This result also implies that \frm{the monopole of $A_{1\mu}$ gauge field in
the corresponding 3+1D $U(1)\times [U(1)\rtimes Z_2]$ SPT state will carries a
projective representation of $U(1)\times [U(1)\rtimes Z_2]$.} Note that the
monopole of $A_{1\mu}$ gauge field does not break the $U(1)\times [U(1)\rtimes
Z_2]$ symmetry.

Again consider only one boundary, we have seen that adding $2\pi$ flux of
$A_{1\mu}$ gauge field
changes the $U(1)\times [U(1)\rtimes Z_2]$ representation of the boundary
excitations from linear to projective.  If the $2\pi$ flux is concentrated
within a region of size $L$, we may assume that the  boundary excitations that
from a  projective representation of  $U(1)\times [U(1)\rtimes Z_2]$ is
concentrated within the region.  When $L$ is large, the $2\pi$ flux is a weak
perturbation.  The fact that a weak perturbation can create an non-trivial
excitation in a projective representation implies that the excitations on the
2+1D boundary $R_t\times M_2'$ is gapless.  To summarize, we have the following
two results: \frm{The 2+1D  $U(1)\times [U(1)\rtimes Z_2]$ gauge theory with
the anomaly described by \eq{fUUZ} is gapless, if we freeze the
$U(1)\times[U(1) \rtimes Z_2]$ gauge fields without break the $U(1)\times
[U(1)\rtimes Z_2]$ symmetry} \frm{The 3+1D  $U(1)\times [U(1)\rtimes Z_2]$ SPT
state characterized by the topological term \eq{fUUZ} of the probe gauge
fields\cite{HW1267,W1375} has gapless boundary excitations, if the $U(1)\times
[U(1)\rtimes Z_2]$ symmetry is not broken.} In other words, the edge of this
particular 3+1D $U(1)\times [U(1)\rtimes Z_2]$ SPT state cannot be a gapped
topologically ordered state that do not break the symmetry.

The first discrete gauge anomaly generates one of the $\Z_2$ in
$H^4[B(U(1)\times [U(1)\rtimes Z_2]),\RZ]= (\RZ)^{\oplus 2}\oplus\Z_2^{\oplus
2}$.  Since Dis$[H^4[B(U(1)\times [U(1)\rtimes Z_2]),\RZ] =\text{Tor}(
\cH^4[U(1)\times [U(1)\rtimes Z_2],\RZ])= \Z_2^{\oplus 2}$, the first discrete
gauge anomaly also  generates one of the $\Z_2$ in Tor$(\cH^4[U(1)\times
[U(1)\rtimes Z_2],\RZ])$. 
According to the K\"unneth formula (see \eqn{kunnU})
\begin{align}
&\ \ \ \
 \cH^4[U(1)\times [U(1)\rtimes Z_2],\RZ]
\nonumber\\
& = 
\cH^2(U(1), \cH^2[U(1)\rtimes Z_2,\RZ]) \oplus
\nonumber\\ & \ \ \ \
\cH^0(U(1), \cH^4[U(1)\rtimes Z_2,\RZ])
\end{align}
where we have only kept the non-zero terms, and
\begin{align}
&\ \ \ \
\cH^2(U(1), \cH^2[U(1)\rtimes Z_2,\RZ]) 
\nonumber\\
&=
\cH^2[U(1), \Z_2] = \Z_2 ,
\end{align}
\begin{align}
&\ \ \ \
\cH^0(U(1), \cH^4[U(1)\rtimes Z_2,\RZ]) 
\nonumber\\
&=
\cH^4[U(1)\rtimes Z_2,\RZ] = \Z_2 .
\end{align}
So the discrete gauge anomaly generates the $\Z_2$ of $\cH^2(U(1),
\cH^2[U(1)\rtimes Z_2,\RZ])$, which is a structure that involve both $U(1)$'s.

\subsubsection{Second discrete gauge anomaly}

In this section, we will discuss the second discrete gauge anomaly that
generates the other $\Z_2$ associated with $\cH^0(U(1), \cH^4[U(1)\rtimes
Z_2,\RZ]) =\cH^4[U(1)\rtimes Z_2,\RZ]$.  The second discrete gauge anomaly is
actually a gauge anomaly of $U(1) \rtimes Z_2$ described by the non-trivial
element in $\cH^4[ U(1)\rtimes Z_2,\RZ]= \Z_2$.  At the moment, we do not
know how to use a 3+1D gauge topological term to describe such an anomaly.
However, we can describe the physical properties (\ie the topological
invariants) of the second discrete gauge anomaly.\cite{W1375}

Let the 3+1D space-time has a topology $R_t\times I\times S_1\times S_1'$.  The
theory on a boundary $R_t\times S_1\times S_1'$ has the second $U(1)\times
[U(1)\rtimes Z_2]$ gauge anomaly.  If we freeze the  $U(1)\times [U(1)\rtimes
Z_2]$ gauge fields without break the $U(1)\times [U(1)\rtimes Z_2]$ symmetry
and consider the large $S_1$ small $S_1'$ limit, then the excitations on $S_1$
are gapped with a non-degenerate ground state, if the $A_{2\mu}$ gauge field is
zero on $S_1\times S_1'$.  However, the excitations on $S_1$ will be gapless or
have degenerate ground states, if there is $\pi$ flux of $A_{2\mu}$ gauge field
going through $S_1'$.\cite{W1375} (The gapless or degenerate ground states on
$S_1$ are edge state of non-trivial 2+1D $Z_2$ SPT state.) Since adding $\pi$
flux to small $S_1'$ is not a small perturbation, we cannot conclude that the
excitations on the 2+1D boundary $R_t\times S_1\times S_1'$ are gapless.

We also note that the  monopole of $A_{2\mu}$ gauge field in the 3+1D bulk
breaks the $Z_2$ symmetry.  In this case, we can only discuss the $U(1)\times
U(1)$ charges of the monopoles (see \Ref{HW1267}).


\section{Understand gauge anomalies through SPT states}
\label{gaSPT}

After discussing some examples of  gauge anomalies, let us turn to the task of
trying to classify gauge anomalies of gauge group $G$.  We will do so by
studying a system with on-site symmetry $G$ in one-higher dimension.  We have
described the general idea of such an approach in section \ref{gen}.  In this
section, we will give more details. 

\subsection{The emergence of non-on-site symmetries in bosonic systems}

Before discussing gauge anomalies, let us introduce the notion of non-on-site
symmetries, and discuss the emergence and a classification of non-on-site
symmetries.  The non-on-site symmetries appear in the low energy boundary
effective theory of a SPT state. So let us first give a brief introduction of
SPT state.

Recently, it was shown that bosonic short-range entangled states\cite{CGW1038}
that do not break any symmetry can be constructed from the elements in group
cohomology class $\cH^{d+1}(G,\R/\Z)$ in $d$ spatial dimensions, where $G$ is
the symmetry group.\cite{CLW1141,CGL1172,CGL1204}  Such symmetric short-range
entangled states are called symmetry-protected trivial (SPT) states or
symmetry-protected topological (SPT) states.

A bosonic SPT state is the ground state of a \emph{local bosonic system} with
an \emph{on-site symmetry} $G$.  A \emph{local bosonic system} is a Hamiltonian
quantum theory with a total Hilbert space that has direct-product structure:
$\cH= \otimes _i \cH_i$ where $\cH_i$ is the local Hilbert space on site-$i$
which has a \emph{finite} dimension.  An \emph{on-site symmetry} is a
representation $U(g)$ of $G$ acting on the total Hilbert space $\cH$ that have
a product form
\begin{align} U(g)= \otimes _i
U_i(g), \ \ \ g \in G, 
\end{align} 
where $U_i(g)$ is a  representation $G$
acting on the local Hilbert space $\cH_i$ on site-$i$.  

A bosonic SPT state is also a \emph{short-range entangled state} that is
invariant under $U(g)$.  The notion of \emph{short-range entangled state} is
introduced in \Ref{CGW1038} as a state that can be transformed into a product
state via a local unitary transformation.\cite{VCL0501,HW0541,V0705} A SPT
state is always a gapped state. It can be smoothly deformed into a gapped
product state via a path that \emph{may} break the symmetry without gap-closing
and phase transitions.  However, a non-trivial SPT state cannot be smoothly
deformed into a gapped product state via any path that \emph{does not} break
the symmetry without phase transitions.

Since SPT states are short-range entangled, it is relatively easy to understand
them systematically. In particular, a systematic construction of the bosonic
SPT state in $d$ spatial dimensions with on-site symmetry $G$ can be obtained
through the group cohomology class
$\cH^{d+1}(G,\R/\Z)$.\cite{CLW1141,CGL1172,CGL1204}

The SPT states are gapped with no ground state degeneracy when there is no
boundary.  If we consider a $d$-space-dimensional bosonic SPT state with a
boundary, then any low energy excitations must be boundary excitations.  Also
since the SPT state is a short-range entangled state, those low energy boundary
excitations can be described by a pure local boundary
theory.\cite{CLW1141,CGL1172,CGL1204} However, if the SPT state is non-trivial
(\ie described by a non-trivial element in $\cH^{d+1}(G,\R/\Z)$), then the
symmetry transformation $G$ must act as a \emph{non-on-site
symmetry}\cite{CW1217,CLW1141,CGL1172,CGL1204} in the effective boundary
theory.  The non-on-site symmetry action $U(g)$ does not have a product form
$U(g)=\otimes_i U_i(g)$.  So the SPT phases in $d$ spatial dimensions lead to
the emergence of non-on-site symmetry in $d-1$ spatial dimensions.  As a
result, the different types of non-on-site symmetry in $(d-1)$ spatial
dimensions are described by $\cH^{d+1}(G,\R/\Z)$.

The non-on-site symmetry has another very interesting (conjectured) property:
\frm{the ground states of a system with a non-on-site symmetry must be
degenerate or gapless.\cite{CLW1141,CGL1172,CGL1204,VS1258,W1375}  The
degeneracy may be due to the symmetry breaking, topological
order,\cite{Wtop,Wrig} or both.}
The above result is proven only in 1+1D.\cite{CLW1141}  For certain types of
non-on-site symmetries, the ground state may even have to be gapless, if the
symmetry is not broken.

For a reason that we will explain later, we will refer non-on-site symmetry as
anomalous symmetry and on-site symmetry as anomaly-free symmetry.  We see
that a system with an anomalous symmetry cannot have a ground state that is
non-degenerate.  
On the other hand a system with an anomaly-free symmetry can have a ground
state that is non-degenerate (and symmetric).  So the anomaly-free property of
a global symmetry is a \emph{sufficient condition} for \emph{the existence of a
gapped ground state that do not break any symmetry.}

\subsection{Anomalous  gauge theories as the boundary effective theory
of bosonic SPT states}

We can alway generalize an on-site global symmetry transformation into a local
gauge transformation by making $g$ to be site dependent \begin{align}
U_\text{gauge}(\{g_i\})=\otimes_i U_i(g_i) \end{align} which is a
representation of $G^{N_s}$, where $N_s$ is the number of sites: \begin{align}
U_\text{gauge}(\{h_i\}) U_\text{gauge}(\{g_i\})= U_\text{gauge}(\{h_ig_i\}).
\end{align} So we say that the on-site symmetry (\ie the anomaly-free symmetry) 
is ``gaugable''.

On the other hand, the non-on-site symmetry of the boundary effective theory is
not ``gaugable''.  If we try to generalize a non-on-site symmetry
transformation to a local gauge transformation: $U_\text{non-on-site}(g)\to
U_\text{gauge}(\{g_i\})$, then $ U_\text{gauge}(\{g_i\})$ does not form a
representation of $G^{N_s}$.  In fact, if we do ``gauge'' the non-on-site
symmetry, we will get an anomalous gauge theory with gauge group $G$ on the
boundary, as demonstrated in \Ref{CGL1172,LV1219,LW1224,CW1217,SL1204} for
$G=U(1),\ SU(2)$.  Therefore, \emph{gauge anomaly $\sim$ non-on-site symmetry}.
This is why we also refer the non-on-site symmetry as anomalous symmetry.
Gauging anomalous symmetry will lead to an anomalous gauge theory. 

Since non-site symmetries emerge at the boundary of SPT states.  Thus gauging
the symmetry in the SPT state in $(d+1)$-dimensional space-time is a systematic
way to construct anomalous gauge theory in  $d$-dimensional space-time.  Then
from the group cohomology description of the SPT states, we find that \emph{the
gauge anomalies in bosonic gauge theories with a gauge group $G$ in $d$
space-time dimensions are described by $\cH^{d+1}(G,\R/\Z)$ (at least
partially).} 

\subsection{The gauge non-invariance (\ie the gauge anomaly) of non-on-site
symmetry and the cocycles in group cohomology}

The standard understanding of gauge anomaly is its ``gauge non-invariance''.
However, in above, we introduce gauge anomaly through SPT state. In this
section, we will show that the two approaches are equivalent.  We also discuss
a direct connection between gauge non-invariance and the group cocycles in
$\cH^{d+1}(G,\R/\Z)$.

The SPT state in the $(d+1)$-dimensional space-time bulk manifold $M$ can be
described by a non-linear $ \sigma $-model with $G$ as the target space
\begin{align} 
S= \int_M \dd^{d+1} x\ \Big[\frac{1}{ \lambda _s} [\prt
g(x^\mu)]^2 + \ii W_\text{top}(g) \Big].  
\end{align} 
in large $ \lambda _s$
limit.  Here we triangulate the $(d+1)$-dimensional bulk manifold $M$ to make it
a (random) lattice or a $(d+1)$-dimensional complex. The field
$g(x^\mu)$ live on the
vertices of the complex.  So $\int \dd^{d+1} x$ is in fact a sum over lattice
sites and $ \partial $ is the lattice difference operator.  The above action
$S$ actually defines a lattice theory.  $\ii W_\text{top}(g)$ is a lattice
topological term which is defined and classified by the elements in
$\cH^{d+1}(G,\R/\Z)$.\cite{CGL1172,CGL1204,LG1220,HW1267,HWW1295,HW1227} This is why
the bosonic SPT states are classified by $\cH^{d+1}(G,\R/\Z)$.

Since $G$ is an on-site symmetry in the $d+1$D bulk, we can always gauge the
on-site symmetry to obtain a gauge theory in the bulk by integrating out
$g(x^\mu)$
\begin{align} 
S= \int \dd^{d+1} x\ \Big[ \frac{\Tr(F_{\mu\nu})^2}{ \lambda_g } +
\ii W^\text{gauge}_\text{top}(A_\mu) \Big].  
\end{align} 
The resulting topological term $ W^\text{gauge}_\text{top}(A)$ in the gauge
theory is always a ``quantized'' topological term discussed in \Ref{HW1267}.
It is a generalization of the Chern-Simons term.\cite{DW9093,HW1267,HWW1295} It is also
related to the topological term $  W_\text{top}(g)$ in the non-linear $ \sigma
$-model when $A_\mu$ is a pure gauge 
\begin{align}
W^\text{gauge}_\text{top}(A_\mu)= W_\text{top}(g), \ \ \ \text{ where }
A_\mu=g^{-1} \partial _\mu g.  
\end{align} 
(A more detailed description of the two topological terms $W_\text{top}(g)$ and
$W^\text{gauge}_\text{top}(A_\mu)$ on lattice can be found in \Ref{HW1267}.) So
the quantized topological term $ W^\text{gauge}_\text{top}(A)$ in the gauge
theory is also described by $\cH^{d+1}(G,\R/\Z)$.

Since $W_\text{top}(g)$ is a cocycle in $\cH^{d+1}(G,\R/\Z)$, 
we have\cite{CGL1172,CGL1204}
\begin{align} 
\th[g(x^\mu)]=
\int_M \dd^{d+1}x\ W^\text{gauge}_\text{top}(g^{-1} \partial _\mu
g) = 0 \text{ mod }2\pi  
\end{align} 
if the space-time $M$ has no boundary.
But if the space-time $M$ has a boundary, then 
\begin{align} 
\th[g(x^\mu)]=
\int_M \dd^{d+1}x\
W^\text{gauge}_\text{top}(g^{-1} \partial _\mu g) \neq 0 \text{ mod }2\pi
\end{align} which represents a gauge non-invariance (or a gauge anomaly) of
the gauged bulk theory in  $(d+1)$-dimensional space-time.  (This is just like
the gauge non-invariance of the Chern-Simons term, which is a special case of
$W^\text{gauge}_\text{top}(A_\mu)$.) 
Note that the gauge  anomaly $\th[g(x^\mu)]$ mod $2\pi$
only depend on $g(x^\mu)$ on the boundary of $M$.
Such a gauge anomaly is canceled by the
boundary theory which is an anomalous bosonic gauge theory.  Such a
point was discussed in detail for $G=U(1),\ SU(2)$ in \Ref{W9125}.

From the above discussion, we see that the bulk theory on the
$(d+1)$-dimensional complex $M$ is gauge invariant if $M$ has no boundary, but
may not be gauge invariant if $M$ has a boundary.  Since a gauge transformation
$g(x^\mu)$ lives on the vertices, it is described by $\{g_i| i \text{ labels
vertices}\}$.  Thus, the gauge non-invariance of the bulk theory is described
by a mapping from $G^{N_s}$ to phase $2\pi \R/\Z$: $ \theta (\{g_i\}_M)$, where
$N_s$ is the number of lattice sites (\ie the number of the vertices).  Such a
mapping has two properties. The first one is
\begin{align} 
\label{lcl}
\theta (\{g_i\}_M) &= \text{ sum of local terms for the cells in } M
\end{align} 
[\ie $\th(\{g_i\}_M)=
\int_M \dd^{d+1}x\ W^\text{gauge}_\text{top}(g^{-1} \partial _\mu g)$].
The second one is
\begin{align} 
\label{th2pi} 
\theta (\{g_i\}_M) &= 0 \text{ mod } 2\pi 
\end{align} 
if $M$ has no boundary, since the theory is gauge invariant when $M$ has no
boundary.  \Eqn{th2pi} is the cocycle condition in group cohomology theory and
the function $\theta (\{g_i\}_M)$ satisfying \eq{th2pi} is a cocycle.

When $M$ does has a boundary, the gauge non-invariance $\theta (\{g_i\}_M)$
only depend on $g_i$'s on the boundary (mod $2\pi$).  So it is a gauge
non-invariance (or a gauge anomaly) on the $d$-dimensional boundary.  Some
times, such a gauge non-invariance $\theta (\{g_i\}_M)$ can be expressed as the
sum of local terms for the cells on the boundary $\prt  M$ [this is potentially
possible since $\theta (\{g_i\}_M)$ only depend on $g_i$'s on the boundary mod
$2\pi$], then such a $\theta (\{g_i\}_M)$ will be called coboundary. The
associated gauge non-invariance is an artifact of us adding gauge non-invariant
boundary terms as we create the boundary of the space-time.  Such a gauge
non-invariance is removable.  So a coboundary does not represent a gauge
anomaly.  Only those gauge non-invariance $\theta (\{g_i\}_M)$ that cannot be
expressed as the sum of local terms represent real  gauge anomalies.  After we
mod out the coboundaries from the cocycles, we obtain $\cH^{d+1}(G,\R/\Z)$.
This way, we see more directly that \frm{the elements in $\cH^{d+1}(G,\R/\Z)$
describe the gauge anomalies in $d$-dimensional space-time for gauge group $G$,
assuming the gauge transformations are described by $\{g_i\}$ on the vertices
of the space-time complex $M$.} We also see that a non-trivial gauge anomaly
(described by a non-trivial cocycle) represents a gauge non-invariance in the
boundary gauge theory.  We believe that the above argument is very general. It
applies to both continuous and discrete gauge groups, and both bosonic theories
and fermionic theories.  (However,  fermionic theories may contain extra
structures. See section \ref{bfGA}.) It turns out that the free part of
$\cH^{d+1}(G,\RZ)$, Free$[\cH^{d+1}(G,\RZ)]$, gives rise to the well known
Adler-Bell-Jackiw anomaly.  The torsion part of $\cH^{d+1}(G,\RZ)$ correspond
to new types of gauge anomalies called nonABJ gauge anomalies.



\section{More general gauge anomalies}

\subsection{$d$-dimensional gauge anomalies and $(d+1)$-dimensional gauge
topological terms}

In the last section, when we discuss the connection between gauge
non-invariance and the group cocycles, we assume that the gauge transformations
on the vertices of the space-time complex $M^d$, $\{g_i\}$, can be arbitrary.
However, in this paper, we want to understand the gauge anomalies in
weak-coupling gauge theories in $d$ space-time dimensions, where gauge field
strength is small.  In this case,  gauge transformations $\{g_i\}$ on the
vertices are not arbitrary.

For finite gauge group $G$, the gauge transformations $\{g_i\}$ on the
vertices of the space-time complex $M^d$ are indeed  arbitrary. So
\frm{$\cH^{d+1}(G,\R/\Z)$ classifies the bosonic gauge anomalies in
$d$-dimensional space-time for finite gauge group $G$.}  
\frm{$\cH^{d+1}(G,\R/\Z)$ partially describes the fermionic gauge anomalies in
$d$-dimensional space-time for finite gauge group $G$.}  
We will discuss the distinction between gauge anomalies in bosonic and
fermionic gauge theory in section \ref{bfGA}.

However, for continuous gauge group $G$, we further require that gauge
transformations $\{g_i\}$  on the vertices of the space-time complex $M^d$ are
close to smooth functions on the space-time manifold.  In this case, there are
more general gauge anomalies.  Free$[\cH^{d+1}(G,\RZ)]$ still describes
all the Adler-Bell-Jackiw  anomaly. But there are nonABJ anomalies that
are beyond  Tor$[\cH^{d+1}(G,\RZ)]$.

To understand more general nonABJ gauge anomalies beyond
Tor$[\cH^{d+1}(G,\RZ)]$, let us view gauge anomalies in $d$-dimensional
space-time as an obstruction to have a non-perturbative definition (\ie a well
defined UV completion) of the gauge theory in the same dimension.  To
understand such an obstruction, let us consider a theory in $(d+1)$-dimensional
space-time where gapped matter fields couple to a gauge theory of gauge group
$G$.  We view of the gauge field as a non-dynamical probe field and only
consider the excitations of the matter fields.  Since the matter fields are
gapped in the bulk, the low energy excitations only live on the boundary and
are described by a boundary low energy effective theory with the non-dynamical
gauge field.  We like to ask, can we define the boundary low energy effective
theory as a pure boundary theory, instead of defining it as a part of
$(d+1)$-dimensional theory?

This question can be answered by considering the induced gauge topological
terms (the terms that do not depend on space-time metrics) in the
$(d+1)$-dimensional theory as we integrate out the gapped mater fields.  There
are two types of the gauge topological terms that can be induced.  The
first type of gauge topological terms has an action amplitude $\ee^{\ii \int_M
\dd^{d+1} x\ W^\text{gauge}_\text{top}(A_\mu)}$ that can change as we change
the gauge field slightly in a local region:
\begin{align}
 \ee^{\ii \int_M \dd^{d+1} x\ W^\text{gauge}_\text{top}(A_\mu+\del A_\mu)}
 \neq \ee^{\ii \int_M \dd^{d+1} x\ W^\text{gauge}_\text{top}(A_\mu)}.
\end{align}
They are classified by Free$[\cH^{d+1}(G,\RZ)]=$
Free$[H^{d+2}(BG,\Z)]$\cite{DW9093,HW1267} and corresponds to the
Adler-Bell-Jackiw  anomalies in $d$-dimensional space-time.  The Chern-Simons
term is an example of this type of topological terms.

The second type of gauge topological terms has an action amplitude that does
not change under any perturbative modifications of the gauge field in a local
region (away from the boundary):
\begin{align}
\label{delA}
 \ee^{\ii \int_M \dd^{d+1} x\ W^\text{gauge}_\text{top}(A_\mu+\del A_\mu)}
 = \ee^{\ii \int_M \dd^{d+1} x\ W^\text{gauge}_\text{top}(A_\mu)}.
\end{align}
$W^\text{gauge}_\text{top}(A_\mu) = \prt_\mu A_{\nu} \prt_\la A_{\ga}
\eps^{\mu\nu\la\ga}$ is an example of such kind of topological terms.  We will
refer the second type of topological terms as locally-null topological terms.
Some of the locally-null topological terms are described by Tor
$[\cH^{d+1}(G,\RZ)]$.\cite{DW9093,HW1267} 

Since $\ee^{\ii \int \dd^{d+1} x\ W^\text{gauge}_\text{top}(A_\mu)}$ does not
change for any perturbative modifications of the gauge field away from the
boundary, one may naively think that it only depends on the fields on the
boundary and write it as a pure boundary term
\begin{align}
 \ee^{\ii \int_M \dd^{d+1} x\ W^\text{gauge}_\text{top}(A_\mu)}
=
 \ee^{\ii \int_{\prt M} \dd^{d} x\ \cL^\text{gauge}_\text{top}(A_\mu)} .
\end{align}
However, the above is not valid in general since $\ee^{\ii \int \dd^{d+1} x\
W^\text{gauge}_\text{top}(A_\mu)}$ does depend on the bulk gauge field away
from the boundary: $\ee^{\ii \int \dd^{d+1} x\
W^\text{gauge}_\text{top}(A_\mu)}$ can change if the modification in the gauge
field away from the boundary cannot be continuously deformed to zero.
In this case, the appearance of the locally-null gauge topological term in
$(d+1)$-dimensions represents an obstruction to view the $(d+1)$-dimensional
theory as a pure $d$-dimensional boundary theory.  This is why we can study
nonABJ gauge anomalies through $(d+1)$-dimensional locally-null gauge
topological terms.  

\subsection{Classifying space and $\pi$-cohomology classes}
\label{more}

To have a systematic description of the locally-null topological terms, let us
use the notion of the classifying space $BG$ for group $G$.  The gauge
configurations (with weak field strength) on the $(d+1)$-dimensional space-time
manifold $M^{d+1}$ can be described by the embeddings of $M^{d+1}$ into $BG$, $
M^{d+1}\to M^{d+1}_{BG} \subset BG$.\cite{DW9093,HW1267}  So we can rewrite our
quantized topological term as a function of the embeddings $M^{d+1}_{BG}$:
\begin{align}
 \int_{M^{d+1}} \dd^{d+1} x W^\text{gauge}_\text{top}(A_\mu) =
S^\text{gauge}_\text{top}(M^{d+1}_{BG})
\end{align}
One way to construct the topological term is to use the topological
$(d+1)$-cocycles $\nu_{d+1} \in H^{d+1}(BG,\RZ)$:
\begin{align}
 S^\text{gauge}_\text{top}(M^{d+1}_{BG})
=2\pi \<\nu_{d+1}, M^{d+1}_{BG}\>
\end{align}
Note that cocycles are cochains, and cochains are defined as linear maps from
cell-complices $M$ to $\RZ$.  $\<\nu_{d+1}, M\>$ denotes such a linear map. As
a part of  definition, $\<\nu_{d+1}, M^{d+1}_{BG}\>$ satisfies the locality
condition
\begin{align}
\<\nu_{d+1}, M^{d+1}_{BG}\>=
 \text{ sum of local terms for the cells in } M ,
\end{align}
which is similar to \eqn{lcl}.

It turns out that the most general locally-null topological terms can be
constructed from $\pi$-cocycles. By definition, a
$(d+1)$-$\pi$-cocycle $\mu_{d+1}$ is a $(d+1)$-cochain that satisfy the
condition
\begin{align}
 \<\mu_{d+1}, M^{d+1}_{BG}\> =\<\mu_{d+1}, N^{d+1}_{BG}\> \text{ mod } 1
\end{align}
if $M^{d+1}_{BG}$ and $N^{d+1}_{BG}$ have no boundaries and $M^{d+1}_{BG}$ and
$N^{d+1}_{BG}$ are homotopic to each other (\ie $M^{d+1}_{BG}$ and $
N^{d+1}_{BG}$ can deform into each other continuously.) As a comparison, a
$(d+1)$-cocycle $\nu_{d+1}$ are $(d+1)$-cochains that satisfy a stronger
condition
\begin{align}
 \<\nu_{d+1}, M^{d+1}_{BG}\> =\<\nu_{d+1}, N^{d+1}_{BG}\> \text{ mod } 1
\end{align}
if $M^{d+1}_{BG}-N^{d+1}_{BG}$ is a boundary of a $(d+2)$-dimensional cell
complex.  

Let us use $\fZ_\pi^{d+1}(BG,\RZ)$ to denote the set of
$(d+1)$-$\pi$-cocycle.  
Clearly, $\fZ_\pi^{d+1}(BG,\RZ)$ contains the set of 
$(d+1)$-cocycles:  $Z^{d+1}(BG,\RZ) \subset \fZ_\pi^{d+1}(BG,\RZ)$,
which in turn contains the set of
$(d+1)$-coboundaries: $B^{d+1}(BG,\RZ) \subset Z^{d+1}(BG,\RZ)$.
The $\pi$-cohomology class
$\fH_\pi^{d+1}(BG,\RZ)$ is defined as
\begin{align}
 \fH_\pi^{d+1}(BG,\RZ)=\fZ_\pi^{d+1}(BG,\RZ)/B^{d+1}(BG,\RZ).
\end{align}
\ie two $\pi$-cocycles are regard as equivalent if they are differ by a
coboundary.  Clearly $\fH_\pi^{d+1}(BG,\RZ)$ contains $H^{d+1}(BG,\RZ)$ as a
subgroup.
\begin{align}
 H^{d+1}(BG,\RZ) &\equiv Z^{d+1}(BG,\RZ)/B^{d+1}(BG,\RZ)
\nonumber\\
& \subset \fH_\pi^{d+1}(BG,\RZ).
\end{align}
However, although in definition, $\fH_\pi^{d+1}(BG,\RZ)$ is more general than
$H^{d+1}(BG,\RZ)$, at the moment, we do not know if $\fH_\pi^{d+1}(BG,\RZ)$ is
strictly larger than $H^{d+1}(BG,\RZ)$. It might be possible that
$\fH_\pi^{d+1}(BG,\RZ) = H^{d+1}(BG,\RZ)$.

Using the $\pi$-cocycles $\mu_{d+1} \in
\fH_\pi^{d+1}(BG,\RZ)$, we can construct generic locally-null
topological terms as
\begin{align}
 S^\text{gauge}_\text{top}(M^{d+1}_{BG})
=2\pi \<\mu_{d+1}, M^{d+1}_{BG}\> .
\end{align}
Thus locally-null topological terms in weak-coupling gauge theories in
$(d+1)$-dimensional space-time are classified by $\fH_\pi^{d+1}(BG,\RZ)$.  Since
the non-locally-null topological terms are classified by
Free$[\cH^{d+1}(G,\RZ)]$, we obtain
\frm{The gauge anomalies in bosonic weak-coupling \\ 
gauge theories with gauge
group $G$ in $d$-dimensional space-time are classified by
Free$[\cH^{d+1}(G,\RZ)]\oplus \fH_\pi^{d+1}(BG,\RZ)$.} 
\frm{The gauge anomalies in fermionic weak-coupling gauge theories with gauge
group $G$ in $d$-dimensional space-time are partially described by
Free$[\cH^{d+1}(G,\RZ)]\oplus \fH_\pi^{d+1}(BG,\RZ)$.} 

As an Abelian group, $\fH_\pi^{d+1}(BG,\RZ)$ may contain $\RZ$, $\Z$, and/or
$\Z_n$.  $\text{Dis}[\fH_\pi^{d+1}(BG,\RZ)]$ is the discrete part of
$\fH_\pi^{d+1}(BG,\RZ)$, which is obtained by dropping the
$\RZ$ parts.  We can show that, for finite group $G$
(see appendix \ref{HH}),
\begin{align}
 \fH_\pi^{d+1}(BG,\RZ)&=\text{Dis}[ \fH_\pi^{d+1}(BG,\RZ)],
\\
 \fH_\pi^{d+1}(BG,\RZ)&= \text{Tor}[\cH^{d+1}(G,\RZ)]
 = \cH^{d+1}(G,\RZ)
.
\nonumber 
\end{align}

\section{Bosonic gauge anomalies and fermionic gauge anomalies}
\label{bfGA}

Why the $\pi$-cohomology theory developed above fails to classify all the
fermionic gauge anomalies?  In this section, we will reveal the reason for this
failure.  Our discussion also suggests that the  $\pi$-cohomology theory may
provide a classification of all bosonic gauge anomalies.

We have been studying gauge anomalies in $d$-dimensional space-time through a
bulk gapped theory in $(d+1)$-dimensional space-time.  The anomalous gauge
theory is defined as the theory on the $d$-dimensional boundary of the
$(d+1)$-dimensional bulk.  In our discussion, we have made the following
assumption.  We first view the gauge field as non-dynamical probe field (\ie
take the gauge coupling to zero).  When the $(d+1)$-dimensional bulk has
several disconnected boundaries, we assume that the total low energy Hilbert
space of the matter fields for all the boundaries is a direct product of the
low energy Hilbert spaces for each connected boundary.  So the  total low
energy Hilbert space of the matter fields can be described by
\emph{independent} matter degrees of freedom on each boundary.  In this case,
when we glue two boundaries together, other boundary will not be affected.
This assumption allows us to use cochains in the classifying space to describe
the low energy effective theory with boundaries.

In the following, we like to argue that the above assumption is valid for
bosonic theories.  This is because when we studied gauge anomalies, we made an
important implicit assumption: we only study \emph{pure} gauge anomalies.  Had
we broken the gauge symmetry, we would be able to have a non-perturbative
definition of the theory in the same dimension.  This implies that the matter
degrees of freedom in the $(d+1)$-dimensional bulk form a short-range entangled
state\cite{CGW1038} with a trivial intrinsic topological order.  For bosonic
systems, short-range entangled bulk state implies that the total Hilbert space
for all the boundaries is a direct product of the Hilbert spaces for each
connected boundary, for any bulk gauge configurations.  This result can be
obtained directly from the canonical form of the bosonic short-range entangled
states suggested in \Ref{CLW1141,CGL1172}.

However, above argument breaks down for fermionic systems, as demonstrated by
the 2+1D $p+\ii p$/$p-\ii p$ fermionic superconductor with $Z_2\times Z_2$
symmetry.  The edge state of the $p+\ii p$/$p-\ii p$ superconductor is
described by \eqn{MajF} which has a $Z_2\times Z_2$ fermionic gauge anomaly.
If we break the $Z_2\times Z_2$ symmetry down to the fermion parity symmetry,
the 1+1D theory \eq{MajF} can indeed be defined on 1D lattice.  Thus the
$p+\ii p$/$p-\ii p$ superconductor has no intrinsic topological order.
However, we do not know the canonical form for such short-range entangled
fermionic state.  The bulk short-range entanglement does not imply that the
total Hilbert space for all the boundaries is a direct product of the Hilbert
spaces for each connected boundary, for any bulk  $Z_2\times Z_2$ gauge
configurations.  We believe this is the reason why the cohomology theory fail
to described all the fermionic gauge anomalies.

\section{The precise relation between gauge anomalies and SPT states}

Despite the very close connection between gauge anomalies and SPT states,
different gauge  anomalies and different  SPT phases do not have a one-to-one
correspondence.

Remember that the gauge anomaly is a property of a low energy weak-coupling
gauge theory. It is the obstruction to have a non-perturbative definition (\ie a
well defined UV completion) of the gauge theory in the same dimension.  While a
SPT phase is a phase of short-range entangled states with a symmetry.

To see the connection between gauge anomalies and SPT phases, we note that the
low energy boundary excitations of a SPT state in $d+1$ space-time dimensions
can always be described by a pure boundary theory, since the bulk SPT states
are  short-range entangled.  However, the on-site symmetry of the  bulk state
must become a non-on-site symmetry on the boundary, if the bulk state has a
non-trivial SPT order.  If we try to gauge the non-on-site symmetry, it will
lead to an anomalous gauge theory in $d$ space-time dimensions.

Every gauge anomaly can be understood this way.  In other words, every gauge
anomaly correspond a SPT state which give rise to a  non-on-site symmetry on
the boundary.  However, some times, two different gauge anomalies may
correspond to two SPT states that can be smoothly connected to each other.  For
example, 3+1D $U(1)$ gauge topological terms $\int \frac{\th}{2! (2\pi)^2}
\prt_\mu A_{\nu} \prt_\la A_{\ga}\eps^{\mu\nu\la\ga} $ gives rise to different
2+1D $U(1)$ gauge anomalies for different values of $\th$ (see section
\ref{U1B3d}).  However, the  $U(1)$ gauge topological terms with different
values of $\th$ correspond to SPT states that can connect to each other without
phase transition.  Thus, the different 2+1D $U(1)$ gauge anomalies correspond
to the same SPT phase.  The gauge anomalies and the SPT phases in one higher
dimension are related by an exact sequence (a many-to-one mapping):
\begin{align*}
&\ \ \ \
\text{$d$-dimensional gauge anomalies of gauge group $G$}
\nonumber\\
& 
\to
\text{$d+1$-dimensional SPT phases of symmetry group $G$}
\nonumber\\
&\to 0
 .
\end{align*}

Using such a relation between  gauge anomalies and SPT phases, we can introduce
the notions of gapless gauge anomalies and gapped  gauge anomalies.  We know
that some SPT states must have gapless boundary excitations if the symmetry is
not broken at the boundary.  We call those  gauge anomalies that map into such
SPT states as ``gapless gauge anomalies''.  We call the  gauge anomalies that
map into the SPT states that can have a gapped boundary states without the
symmetry breaking ``gapped gauge anomalies''.

It appears that all the ABJ anomalies are gapless gauge anomalies.  The 2+1D
continuous $U(1)$ gauge anomalies discussed above (see section \ref{U1B3d}) are
examples of gapped  gauge anomalies, which are nonABJ anomalies.  The first
discrete 2+1D $U(1)\times (U(1)\rtimes Z_2)$ gauge anomaly discussed in section
\ref{U1U1Z2} is an example of gapless  gauge anomaly, which is also a nonABJ
anomaly.  All the 1+1D gauge anomalies are gapless  gauge anomalies, since 2+1D
SPT state always have gapless edge excitations if the symmetry is not
broken.\cite{CLW1141}

\section{Non-perturbative definition of chiral gauge theories}
 
In this section, we will discuss an application of the deeper understanding of
gauge anomalies discussed in this paper: a lattice non-perturbative definition
of any anomaly-free chiral gauge theories.  This idea can be used to construct
a lattice non-perturbative definition of the $SO(10)$ grant unification chiral
gauge theory.\cite{W1345}

\subsection{Introduction}

The $U(1) \times SU(2)\times SU(3)$  standard
model\cite{G6179,W6764,SW6468,G6267,Z6424,FG7235} is the theory which is
believed to describe all elementary particles (except the gravitons) in nature.
The standard model is a chiral gauge theory where the $SU(2)$ gauge fields
couple differently to right-/left-hand fermions.  For a long time, we only know
a  perturbative definition of the standard model via the perturbative expansion
of the gauge coupling constant.  The perturbative definition is not self
consistent since the perturbative expansion is known to diverge.  In this
section, we would like propose a non-perturbative definition of any
anomaly-free chiral gauge theories. 
We will construct well-regulated Hamiltonian quantum models\footnote{A
``Hamiltonian quantum theory'' is a quantum theory with (1) a finite
dimensional Hilbert space to describe all the states, (2) a Hamiltonian
operator to describe the time evolution of the states, and (3) operators to
describe the physical quantities.} whose low energy effective theory is any
anomaly-free chiral gauge theory.  Our approach will apply to the standard
model if the standard model is free of \emph{all} anomalies.

There are many previous researches that try to give chiral gauge theories a
non-perturbative definition. There are lattice gauge theory
approaches,\cite{K7959} which fail since they cannot reproduce chiral couplings
between the gauge field and the fermions.  There are domain-wall fermion
approaches.\cite{K9242,S9390} But the gauge fields in the domain-wall fermion
approaches propergate in one higher dimension: 4+1 dimensions.  There are also
overlap-fermion approaches.\cite{L9995,N0103,S9947,L0128} However, the
path-integral in overlap-fermion approaches may not describe a  Hamiltonian
quantum theory (for example, the total Hilbert space in the overlap-fermion
approaches, if exist, may not have a finite dimension, even for a space-lattice
of a finite size).

Our construction has a similar starting point as the mirror fermion approach
discussed in \Ref{EP8679,M9259,BMP0628,GP0776}.  However, later work either
fail to demonstrate \cite{GPR9396,L9418,CGP1247} or argue that it is almost
impossible\cite{BD9216} to use  mirror fermion approach to non-perturbatively
define anomaly-free chiral gauge theories.  Here, we will argue that  the
mirror fermion approach actually works.  We are able to use the defining
connection between the chiral gauge theories in $d$-dimensional space-time and
the SPT states in $(d+1)$-dimensional space-time to show that, if a chiral
gauge theory is free of \emph{all} the anomalies, then we can construct a
lattice gauge theory whose low energy effective theory reproduces the
anomaly-free chiral gauge theory.  We show that lattice gauge theory approaches
actually can define anomaly-free chiral gauge theories non-perturbatively
without going to one higher dimension, if we include a proper direct
interactions between lattice fermions.

\subsection{A non-perturbative definition of any 
anomaly-free chiral gauge theories}

\begin{figure}[tb] 
\begin{center} 
\includegraphics[scale=0.42]{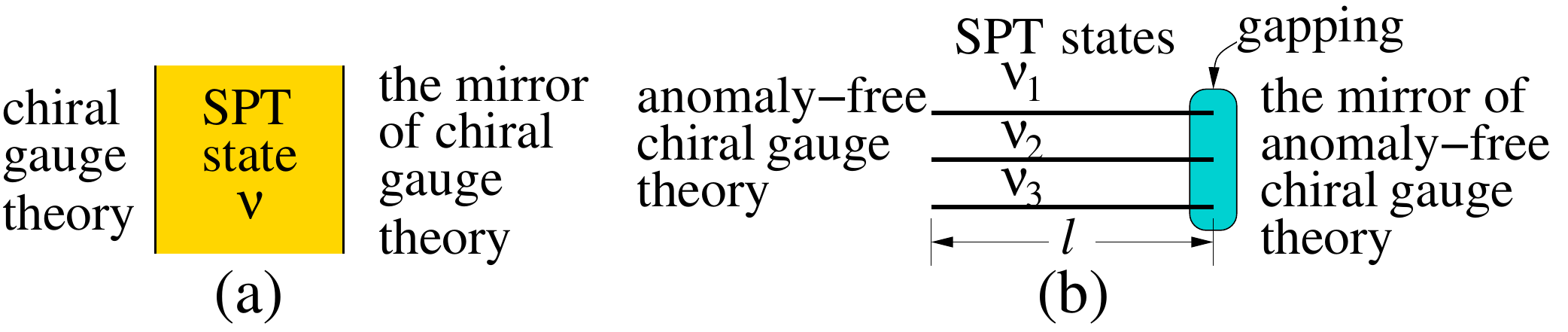}
\end{center}
\caption{ (Color online) (a) A SPT state described by a cocycle $\nu \in
\cH^{d+1}(G,\R/\Z)$ in $(d+1)$-dimensional space-time.  After ``gauging'' the
on-site symmetry $G$, we get a bosonic chiral gauge theory on one boundary and
the ``mirror'' of the  bosonic chiral gauge theory on the other boundary.  (b)
A stacking of a few SPT states in $(d+1)$-dimensional space-time described by
cocycles $\nu_i$.  If $\sum_i \nu_i=0$, then after ``gauging'' the on-site
symmetry $G$, we get a anomaly-free chiral gauge theory on one boundary. We
also get the ``mirror'' of the  anomaly-free chiral gauge theory on the other
boundary, which can be gapped without breaking the ``gauge symmetry''.  }
\label{chgauge} 
\end{figure}

Let us start with a SPT state in $(d+1)$-dimensional space-time with a
on-site symmetry $G$ (see Fig. \ref{chgauge}a). We assume that the SPT
state is described by a cocycle $\nu \in \cH^{d+1}(G,\R/\Z)$.  On the
$d$-dimensional boundary, the low energy effective theory will have a
non-on-site symmetry (\ie an anomalous symmetry) $G$.  Here we will assume that
the $d$-dimensional boundary excitations are gapless and do not break the
symmetry $G$.  After ``gauging'' the on-site symmetry $G$ in the
$(d+1)$-dimensional bulk, we get a bosonic chiral gauge theory on the
$d$-dimensional boundary whose anomaly is described by the cocycle $\nu$. 

Then let us consider a stacking of a few bosonic SPT states in
$(d+1)$-dimensional space-time described by cocycles $\nu_i \in
\cH^{d+1}(G,\R/\Z)$ where the interaction between the SPT states are weak (see
Fig. \ref{chgauge}b).  We also assume that $\sum_i \nu_i=0$.  Because the
stacked system has a trivial SPT order, if we turn on a proper
\emph{$G$-symmetric} interaction between different layers on one of the two
boundaries, we can fully gap the boundary excitations in such a way that the
ground state is not degenerate.  (Such a gapping process also do not break the
$G$ symmetry.) Thus the gapping process does not leave behind any low energy
degrees of freedom on the gapped boundary.  Now we ``gauge'' the on-site
symmetry $G$ in the $(d+1)$-dimensional bulk.  The resulting system is a
non-perturbative definition of anomaly-free bosonic chiral gauge theory
described by $\nu_i$ with $\sum \nu_i=0$.  Since the thickness $l$ of the
$(d+1)$-dimensional bulk is finite (although $l$ can be large so that the two
boundaries are nearly decoupled), the system actually has a $d$-dimensional
space-time.  In particular, due to the finite $l$, the gapless gauge bosons of
the gauge group $G$ are gapless excitations on the $d$-dimensional space-time.

The same approach also works for fermionic systems.  We can start with a few
fermionic SPT states in $(d+1)$-dimensional space-time described by
super-cocycles $\nu_i$\cite{GW1248} that satisfy $\sum \nu_i=0$ (\ie the
combined fermion system is free of all the gauge anomalies).  If we turn on a
proper $G$-symmetric interaction on one boundary, we can fully gap the boundary
excitations in such a way that the ground state is not degenerate and does
break the symmetry $G$.  In this case, if we gauge the bulk on-site symmetry,
we will get a non-perturbative definition of anomaly-free fermionic chiral
gauge theory.


\subsection{A non-perturbative definition of some anomalous chiral gauge
theories}

In the above non-perturbative definition of some anomaly-free chiral gauge
theories, the lattice gauge theories reproduce all the low energy properties of
the anomaly-free chiral gauge theories, including all the low energy
particle-like excitations \emph{and} degenerate ground states.  This is because
the gapped mirror sector on the other boundary has a non-degenerate ground
state.

However, for the application to high energy physics, in particular, for the
application to non-perturbatively define the standard model, we only need the
non-perturbatively defined theory to reproduce  all the low energy
particle-like excitations.  In this case, the gapped mirror sector on the other
boundary can have degenerate ground states and non-trivial topological orders.

If we only need the non-perturbatively defined theory to reproduce  all the low
energy particle-like excitations, we can even define certain anomalous chiral
gauge theories non-perturbatively, following the method outlined in the
previous section.  Using the notions of ``gapless gauge anomalies'' and
``gapped gauge anomalies'' introduced in the last section, we see that we can
use a lattice gauge theory to give non-parturbative definition of an anomalous
chiral gauge theory, if the chiral gauge theory has a ``gapped gauge anomaly''.

Thus all the chiral gauge theories with the ABJ anomalies do not have a
non-perturbative difinition.  The 2+1D chiral gauge theories with the first
discrete 2+1D $U(1)\times (U(1)\rtimes Z_2)$ gauge anomaly discussed in section
\ref{U1U1Z2} also do not have a non-perturbative difinition.  However, many
other anomalous chiral gauge theories have ``gapped gauge anomalies'' and they
do have a non-perturbative difinition.  The gapped boundary states of those
anomalous chiral gauge theories have non-trivial topological orders and ground
state degeneracies.


\section{Summary}

In this paper, we introduced a $\pi$-cohomology theory to systematically
describe gauge anomalies. We propose that bosonic gauge anomalies in
$d$-dimensional space-time for gauge group $G$ are classified by the elements
in Free$[\cH^{d+1}(G,\RZ)]\oplus \fH_\pi^{d+1}(BG,\RZ)$, where
$\fH_\pi^{d+1}(BG,\RZ)$ is the $\pi$-cohomology class of the classifying space
$BG$ of  group $G$.  We show that  the $\pi$-cohomology class
$\fH_\pi^{d+1}(BG,\RZ)$ contains the topological cohomology class
$H_\pi^{d+1}(BG,\RZ)$ as a subgroup.  

The $\pi$-cohomology theory also apply to fermion systems,
where Free$[\cH^{d+1}(G,\RZ)]\oplus \fH_\pi^{d+1}(BG,\RZ)$ describes some of
the fermionic gauge anomalies.  The gauge anomalies for both continuous and
discrete groups are treated at the same footing.

Motivated by the  $\pi$-cohomology theory and the closely related group
cohomology theory, we studied many examples of nonABJ anomalies.  Many results
are obtained, which are stressed by the framed boxes.

The close relation between gauge anomalies and SPT states in one-higher
dimension allows us to give a non-perturbative definition of any anomaly-free
chiral gauge theory in terms of lattice gauge theories.
In this paper, we outline a generic construction to
obtain such a non-perturbative definition.

The close relation between gauge anomalies and SPT states also allows us to
gain a deeper understanding for both gauge anomalies and SPT states.  Such a
deeper understanding suggests that gravitational anomalies are classified by
topological orders\cite{Wtop,Wrig} (\ie patterns of long-range
entanglement\cite{CGW1038}) in one-higher dimension.  To see such a connection,
we like to point out that if a theory cannot be non-perturbatively defined in
the same dimension even after we break all the gauge symmetries, then the
theory should have an anomaly that is beyond the gauge anomaly. This more
general anomaly can be identified as gravitational anomaly.  A theory with
gravitational anomaly can only appear as an effective theory on the boundary of
a bulk theory in one-higher dimension, which has a non-trivial intrinsic
topological order.\cite{Wtop,Wrig}  This line of thinking suggests that the
gravitational anomalies are classified by topological orders (\ie patterns of
long-range entanglement\cite{CGW1038}) in one-higher dimension, leading to a
new fresh point of view on gravitational anomalies.

We also like to remark that in \Ref{HW1267}, quantized topological terms in
$d$-space-time-dimensional weak-coupling gauge theory are systematically
constructed using the elements in $\cH^{d+1}(G,\Z)$.  The study in this paper
shows that more general  quantized topological terms can be constructed  using
the discrete elements in Free$[\cH^{d}(G,\RZ)]\oplus \fH_\pi^{d}(BG,\RZ)$.  

I like to thank Xie Chen, Zheng-Cheng Gu, and Ashvin Vishwanath for many
helpful discussions.  This research is supported by NSF Grant No. DMR-1005541,
NSFC 11074140, and NSFC 11274192.  Research at Perimeter Institute is supported
by the Government of Canada through Industry Canada and by the Province of
Ontario through the Ministry of Research.


\appendix

\section{The nonABJ gauge anomalies and
the global gauge anomalies}

The nonABJ gauge anomalies described by $\fH_\pi^{d+1}(BG,\RZ)$ is closely
related to bosonic global gauge anomalies.  The definition of bosonic global
gauge anomalies is very similar to the definition of the fermionic $SU(2)$
global gauge anomaly first introduced by Witten.\cite{W8224}  In this section,
we will follow Witten's idea to give a definitions of bosonic global gauge
anomalies for \emph{continuous} gauge groups.\cite{CP9045}  We then discuss the
relation between the nonABJ gauge anomalies and newly defined bosonic global
gauge anomalies, for the case of continuous gauge groups.

We like to point out that the bosonic global gauge anomalies defined here
are potential global gauge anomalies. They may or may not be realizable
by boson systems.

\subsection{A definition of bosonic/fermionic  global gauge anomalies for
continuous gauge groups}

\begin{table}[t]
 \centering
 \begin{tabular}{ |c||c|c|c|c|c|c| }
 \hline
 $\pi_d: G\backslash d$ & 1 & 2 & 3 & 4 & 5 & 6  \\
\hline
\hline
$U(1)$ &$\Z$& 0 &   0 &  0 & 0 & 0  \\ 
\hline
$SU(2)$ & 0 & 0 & $\Z$ & $\Z_2$ & $\Z_2$ & $\Z_{12}$ \\ 
$SU(3)$ & 0 & 0 & $\Z$ & 0 & $\Z$ & $\Z_6$ \\ 
$SU(5)$ & 0 & 0 & $\Z$ & 0 & $\Z$ & 0 \\ 
\hline
$SO(3)$ & $\Z_2$ & 0 & $\Z$ & $\Z_2$ & $\Z_2$ & $\Z_{12}$ \\ 
$SO(10)$ & $\Z_2$ & 0 & $\Z$ & 0 & $0$ & 0 \\ 
$Spin(10)$ & 0 & 0 & $\Z$ & 0 & $0$ & 0 \\ 
\hline
 \end{tabular}
 \caption{
A list of homotopy groups $\pi_d(G)$ which describe the global gauge anomalies
in $d$ space-time dimensions.
}
 \label{tb}
\end{table}

We use the gauge non-invariance of the partition function under the ``large''
gauge transformations to define the global gauge anomalies.  Let us consider a
weak-coupling gauge theory in closed $d$-dimensional space-time $S_d$ which has
a spherical topology. We also assume a continuous gauge group $G$.  
If $\pi_{d}(G)$ is non-trivial, it means that there are exist non-trivial
``large'' gauge transformations that does not connect to the identity gauge
transformation (\ie the trivial   gauge transformation).  Note that
$\pi_{d}(G)$ forms a group.  Under a ``large'' gauge transformation, the
partition function may change a phase 
\begin{align} 
Z[A_\mu'] =\ee^{\ii \th}
Z[A_\mu],\ A_\mu'= g^{-1} A_\mu g -\ii g^{-1} \prt_\mu g, 
\end{align} 
where $g(x)$ is a non-trivial map from $M_d$ to $G$.  The different  choices of
the phases $\ee^{\ii \th}$ correspond to different 1D representations of
$\pi_{d}(G)$ which are classified by first group cohomology classes
$\cH^1[\pi_{d}(G),\RZ]$.  So the potential global gauge anomalies are
described by $ \cH^1[\pi_{d}(G),\RZ]$.\\[1mm]  
\frm{The potential global gauge
anomalies in $d$-dimensional space-time and for gauge group $G$ are described
by $\cH^1[\pi_{d}(G),\RZ]$.} 
Since $\pi_d(G)$ is an Abelian group, we have
$\cH^1[\pi_{d}(G),\RZ]=\pi_{d}(G)$.  In table \ref{tb}, we list $\pi_{d}(G)$
for some groups. For a more general discussion of global gauge anomalies
along this line of thinking, see \Ref{CP9045}.

We will refer those global gauge anomalies that appear in a pure bosonic
systems as bosonic global gauge anomalies.  We will refer those global gauge
anomalies that appear in a fermionic systems as fermionic global gauge
anomalies.  Witten's $SU(2)$ global anomaly is a special case of fermionic
global gauge anomalies, which exists because $\pi_4(SU(2))=\Z_2$.  So for a
fermionic $SU(2)$ gauge theory defined on space-time manifold $S_4$, its
partition function $Z[A_\mu]$ may change sign as we make a large $SU(2)$ gauge
transformation:
\begin{align}
 Z[A'_\mu] =-Z[A_\mu],\ 
A_\mu'= g^{-1} A_\mu g -\ii g^{-1} \prt_\mu g,\  g(x) \in G,
\end{align}
where $g(x)$ is a non-trivial map from $S_4$ to $SU(2)$.
This is described by the non-trivial element in $\cH^1[\pi_{4}[SU(2)],\RZ]$.

\subsection{The nonABJ gauge anomalies and
the bosonic global gauge anomalies}

We note that $\pi_d(G)$ also describes the classes of $G$ gauge configurations
on $S_{d+1}$ that cannot be continuously deformed into each others.  Those
classes of $G$ gauge configurations on $S_{d+1}$ correspond to classes of
embedding $S_{d+1} \to BG$ that  cannot be continuously deformed into each
others.  This picture explains a mathematical result $\pi_d(G)=\pi_{d+1}(BG)$.
So the potential global gauge anomalies in $d$-dimensional space-time are
defined as 1D representations $\cH^1[\pi_{d+1}(BG),\RZ]$.  Each
$\pi$-cocycle $\mu_{d+1}$ in $\fH^{d+1}_\pi(BG,\RZ)$ induces an 1D
representations of $\pi_{d+1}(BG)$ via
\begin{align}
\<\mu_{d+1}, S^{d+1}_{BG}\> \text{ mod } 1,
\end{align}
where $S^{d+1}_{BG}$ is an embedding $S_{d+1} \to BG$.
Thus we have a map
\begin{align}
 \fH^{d+1}_\pi(BG,\RZ) \to \cH^1[\pi_{d+1}(BG),\RZ] .
\end{align}
The above map represents the relation between the nonABJ gauge anomalies
described by $\fH^{d+1}_\pi(BG,\RZ)$ and the global gauge anomalies described
by $\cH^{1}(\pi_{d+1}(BG),\RZ)$.  If a 1D representation of $\pi_{d+1}(BG)$
cannot be induced by any $\pi$-cocycle, then the corresponding global
gauge anomaly is not realizable by local bosonic systems.

\section{$\fH^d_\pi(BG,\RZ)= \cH^d(G,\RZ)$ for finite groups}

When $G$ is finite, any closed complex $M_{BG}$ in $BG$ can be deformed
continuously into a canonical form where all the vertices of  $M_{BG}$ is on
the same point in $BG$.  All the edges of $M_{BG}$ is mapped to $\pi_1(BG)=G$.
So each edge of $M_{BG}$ is labeled by a group element.  All the  canonical
complex $M_{BG}$, with all the vertices on the same point and with  fixed the
group elements on all the edges, can deform into each other, since $\pi_n(BG)
=0$ for $n>1$ if $G$ is finite.  In this case, an evaluation of a $\pi$-cocycle
on $M_{BG}$ is a function of the group elements on the edges.  Such a function
is a group cocycle. This way we map a  $\pi$-cocycle to a group cocycle.

We also note that the group cocycle condition implies that the evaluation on
any $d$-sphere is trivial. So a group cocycle is also a $\pi$-cocycle.  The fact
that $\pi$-cocycle = group cocycle for finite groups allows us to show
$\fH^d_\pi(BG,\RZ)= \cH^d(G,\RZ)$.

\section{Relation between $H^{d+1}(BG,\Z)$ and $\cH_B^d(G,\R/\Z)$} 
\label{HBGZHcGRZ}

We can show that the topological cohomology of the classifying space,
$H^{d+1}(BG,\Z)$, and the Borel-group cohomology, $\cH_B^d(G,\R/\Z)$, are
directly related 
\begin{align}
\label{HdHd1b}
   H^{d+1}(BG,\Z) \simeq  \cH_B^d(G,\R/\Z) .
\end{align}
This result is obtained from \Ref{WW1104}.  On page 16 of \Ref{WW1104}, it is mentioned
in Remark IV.16(3) that $\cH^d_B(G,\R)= \Z_1$ (there, $\cH^d_B(G,M)$ is denoted
as $\cH^d_\text{Moore}(G,M)$ which is equal to $\cH^d_\text{SM}(G,M)$).  It is
also shown in Remark IV.16(1) and in Remark IV.16(3) that
$\cH^d_\text{SM}(G,\Z)=H^{d}(BG,\Z)$ and
$\cH^d_\text{SM}(G,\R/\Z)=H^{d+1}(BG,\Z)$,
(where $G$ can have a non-trivial action on $\R/\Z$ and $\Z$, and
$H^{d+1}(BG,\Z)$ is the usual topological cohomology on the classifying space
$BG$ of $G$).
Therefore, 
we have
\begin{align} 
\label{HdR} 
& \cH^d_B(G,\R/\Z)=\cH^{d+1}_B(G,\Z)=H^{d+1}(BG,\Z),
\nonumber\\
& \cH^d_B(G,\R)=\Z_1,\ \ d>0.  
\end{align} 
These results are valid for both continuous groups and discrete groups, as well
as for $G$ having a non-trivial action on the modules $\R/\Z$ and $\Z$.  

\section{Group cohomology $\cH_B^*(G,\M)$ and topological
cohomology $H^*(BG,\M)$ on the classifying space }
\label{HH}

First, we can show that
\begin{align}
\label{HdHd1b}
   H^{d+1}(BG,\Z) \simeq  \cH_B^d(G,\R/\Z) .
\end{align}
where $\cH_B^d(G,\R/\Z)$ is the Borel group cohomology classes.  In the main
text of this paper, we drop the subscript $B$.  This result is obtained from
\Ref{WW1104}.  On page 16 of \Ref{WW1104}, it is mentioned in Remark IV.16(3)
that $\cH^d_B(G,\R)= 0$ (there, $\cH^d_B(G,\M)$ is denoted as
$\cH^d_\text{Moore}(G,\M)$ which is equal to $\cH^d_\text{SM}(G,\M)$).  It is
also shown in Remark IV.16(1) and in Remark IV.16(3) that
$\cH^d_\text{SM}(G,\Z)=H^{d}(BG,\Z)$ and
$\cH^d_\text{SM}(G,\R/\Z)=H^{d+1}(BG,\Z)$,
(where $G$ can have a non-trivial action on $\R/\Z$ and $\Z$, and
$H^{d+1}(BG,\Z)$ is the usual topological cohomology on the classifying space
$BG$ of $G$).
Therefore, 
we have
\begin{align} 
\label{HdR} 
& \cH^d_B(G,\R/\Z)=\cH^{d+1}_B(G,\Z)=H^{d+1}(BG,\Z),
\nonumber\\
& \cH^d_B(G,\R)=0,\ \ d>0.  
\end{align} 
These results are valid for both continuous groups and discrete groups, as well
as for $G$ having a non-trivial action on the modules $\R/\Z$ and $\Z$.  We see
that, for integer coefficient, $\cH_B^d(G,\Z)$ and $H^d(BG,\Z)$ are the same.

To see how $\cH_B^d(G,\RZ)$ and $H^d(BG,\RZ)$ are related, we can use the
universal coefficient theorem \eq{ucf} to compute $H^d(BG,\RZ)$:
\begin{align}
& H^d(BG,\RZ)=\text{Con}[ H^d(BG,\Z) ]\oplus
\text{Tor}[ H^{d+1}(BG,\Z) ]
\nonumber\\
& =\text{Con}[ \cH_B^d(G,\Z) ]\oplus
\text{Tor}[ \cH_B^{d+1}(G,\Z) ]
\nonumber\\
& =\text{Con}[ \cH_B^{d-1}(G,\RZ) ]\oplus
\text{Tor}[ \cH_B^{d}(G,\RZ) ]
,
\end{align} 
where $\text{Con}[\Z]=\RZ$, $\text{Con}[\Z_n]=0$, and
$\text{Con}[\M_1\oplus \M_2]= \text{Con}[\M_1]\oplus  \text{Con}[\M_2]$. 

For $d =$ odd, we also have
\begin{align}
 \text{Free}[H^d(BG,\Z)] &=
 \text{Free}[\cH_B^{d-1}(G,\RZ)] =0,
\nonumber\\
 H^d(BG,\RZ) & =\text{Tor}[H^{d+1}(BG,\Z)].
\nonumber\\
&= \text{Tor}[\cH_B^{d}(G,\RZ)]
\end{align}

For finite group $G$ and any $d$, we have
\begin{align}
 \text{Free}[H^d(BG,\Z)] &=
 \text{Free}[\cH_B^{d-1}(G,\RZ)] =0,
\nonumber\\
 H^d(BG,\RZ) & =\text{Tor}[H^{d+1}(BG,\Z)].
\nonumber\\
&= \text{Tor}[\cH_B^{d}(G,\RZ)]
\end{align}

\section{The K\"unneth formula} \label{KF}

The K\"unneth formula is a very helpful formula that allows us to calculate the
cohomology of chain complex $X\times X'$ in terms of  the cohomology of chain
complex $X$ and chain complex $X'$.  The K\"unneth formula is given by (see
\Ref{Spa66} page 247)
\begin{align}
\label{kunn}
&\ \ \ \ H^d(X\times X',\M\otimes_R \M')
\nonumber\\
&\simeq \Big[\oplus_{k=0}^d H^k(X,\M)\otimes_R H^{d-k}(X',\M')\Big]\oplus
\nonumber\\
&\ \ \ \ \ \
\Big[\oplus_{k=0}^{d+1}
\text{Tor}_1^R(H^k(X,\M),H^{d-k+1}(X',\M'))\Big]  .
\end{align}
Here $R$ is a principle ideal domain and $\M,\M'$ are $R$-modules such that
$\text{Tor}_1^R(\M,\M')=0$.  We also require that $\M'$ and $H^d(X',\Z)$ are
finitely generated, such as $\M'=\Z\oplus \cdots \oplus \Z\oplus
\Z_n\oplus\Z_m\oplus\cdots$.  

A $R$-module is like a vector space over $R$ (\ie we can ``multiply'' a vector
by an element of $R$.) For more details on principal ideal domain and
$R$-module, see the corresponding Wiki articles.  Note that $\Z$ and $\R$ are
principal ideal domains, while $\R/\Z$ is not.  Also, $\R$ and $\R/\Z$ are not
finitely generate $R$-modules if $R=\Z$.  The K\"unneth formula works for
topological cohomology where $X$ and $X'$ are treated as topological spaces.
The K\"unneth formula also works for group cohomology, where $X$ and $X'$ are
treated as groups, $X=G$ and $X'=G'$, provided that $G'$ is a finite group.
However, the above K\"unneth formula  does not apply for Borel-group cohomology
when $X'=G'$ is a continuous group, since in that case $\cH_B^d(G',\Z)$ is not
finitely generated.

The tensor-product operation $\otimes_R$ and the torsion-product operation
$\text{Tor}_1^R$ have the following properties:
\begin{align}
\label{tnprd}
& A \otimes_\Z B \simeq B \otimes_\Z A ,
\nonumber\\
& \Z \otimes_\Z \M \simeq \M \otimes_\Z \Z =\M ,
\nonumber\\
& \Z_n \otimes_\Z \M \simeq \M \otimes_\Z \Z_n = \M/n\M ,
\nonumber\\
& \Z_n \otimes_\Z \RZ \simeq \RZ \otimes_\Z \Z_n = 0,
\nonumber\\
& \Z_m \otimes_\Z \Z_n  =\Z_{\<m,n\>} ,
\nonumber\\
&  (A\oplus B)\otimes_R \M = (A \otimes_R \M)\oplus (B \otimes_R \M)   ,
\nonumber\\
& \M \otimes_R (A\oplus B) = (\M \otimes_R A)\oplus (\M \otimes_R B)   ;
\end{align}
and
\begin{align}
\label{trprd}
& \text{Tor}_1^R(A,B) \simeq \text{Tor}_1^R(B,A)  ,
\nonumber\\
& \text{Tor}_1^\Z(\Z, \M) = \text{Tor}_1^\Z(\M, \Z) = 0,
\nonumber\\
& \text{Tor}_1^\Z(\Z_n, \M) = \{m\in \M| nm=0\},
\nonumber\\
& \text{Tor}_1^\Z(\Z_n, \RZ) = \Z_n,
\nonumber\\
& \text{Tor}_1^\Z(\Z_m, \Z_n) = \Z_{\<m,n\>} ,
\nonumber\\
& \text{Tor}_1^R(A\oplus B,\M) = \text{Tor}_1^R(A, \M)\oplus\text{Tor}_1^R(B, \M),
\nonumber\\
& \text{Tor}_1^R(\M,A\oplus B) = \text{Tor}_1^R(\M,A)\oplus\text{Tor}_1^R(\M,B)
,
\end{align}
where $\<m,n\>$ is the greatest common divisor of $m$ and $n$.  These
expressions allow us to compute the tensor-product $\otimes_R$ and  the
torsion-product $\text{Tor}_1^R$.


As the first application of K\"unneth formula, we like to use it to calculate
$H^*(X',\M)$ from $H^*(X',\Z)$,  by choosing $R=\M'=\Z$. In this case, the
condition $\text{Tor}_1^R(\M,\M')=\text{Tor}_1^{\Z}(\M,\Z)=0$ is always
satisfied. So we have
\begin{align}
\label{kunnZ}
&\ \ \ \ H^d(X\times X',\M)
\nonumber\\
&\simeq \Big[\oplus_{k=0}^d H^k(X,\M)\otimes_{\Z} H^{d-k}(X',\Z)\Big]\oplus
\nonumber\\
&\ \ \ \ \ \
\Big[\oplus_{k=0}^{d+1}
\text{Tor}_1^{\Z}(H^k(X,{\M}),H^{d-k+1}(X',\Z))\Big]  .
\end{align}
The above is valid for topological cohomology.
It is also valid for group  cohomology:
\begin{align}
\label{kunnZG1}
&\ \ \ \ \cH^d(G\times G',\M)
\nonumber\\
&\simeq \Big[\oplus_{k=0}^d \cH^k(G,\M)\otimes_{\Z} \cH^{d-k}(G',\Z)\Big]\oplus
\nonumber\\
&\ \ \ \ \ \
\Big[\oplus_{k=0}^{d+1}
\text{Tor}_1^{\Z}(\cH^k(G,\M),\cH^{d-k+1}(G',\Z))\Big]  .
\end{align}
provided that $G'$ is a finite group.
Using \eqn{HdR}, we can rewrite the above as
\begin{align}
\label{kunnZG2}
&\ \ \ \ \cH^d(G\times G',\M)
\simeq 
\cH^d(G,\M)\oplus 
\nonumber\\
&\ \ \ \ \ \
\Big[\oplus_{k=0}^{d-2} \cH^k(G,\M)\otimes_{\Z} \cH^{d-k-1}(G',\RZ)\Big]\oplus
\nonumber\\
&\ \ \ \ \ \
\Big[\oplus_{k=0}^{d-1}
\text{Tor}_1^{\Z}(\cH^k(G,\M),\cH^{d-k}(G',\RZ))\Big]  ,
\end{align}
where we have used
\begin{align}
 \cH^1(G',\Z)=0.
\end{align}
If we further choose $\M=\RZ$, we obtain
\begin{align}
\label{kunnZG}
&\ \ \ \ \cH^d(G\times G',\RZ)
\nonumber\\
&\simeq 
\cH^d(G,\RZ)\oplus 
\cH^d(G',\RZ)\oplus 
\nonumber\\
&\ \ \ \ \ \
\Big[\oplus_{k=1}^{d-2} \cH^k(G,\RZ)\otimes_{\Z} \cH^{d-k-1}(G',\RZ)\Big]\oplus
\nonumber\\
&\ \ \ \ \ \
\Big[\oplus_{k=1}^{d-1}
\text{Tor}_1^{\Z}(\cH^k(G,\RZ),\cH^{d-k}(G',\RZ))\Big]  ,
\end{align}
where $G'$ is a finite group.

We can further choose $X$ to be the space of one point (or the trivial group of
one element) in \eqn{kunnZ} or \eqn{kunnZG1}, and use
\begin{align}
H^{d}(X,\M))=
\begin{cases}
\M, & \text{ if } d=0,\\
0, & \text{ if } d>0,
\end{cases}
\end{align}
to reduce \eqn{kunnZ} to
\begin{align}
\label{ucf}
 H^d(X,\M)
&\simeq  \M \otimes_{\Z} H^d(X,\Z)
\oplus
\text{Tor}_1^{\Z}(\M,H^{d+1}(X,\Z))  .
\end{align}
where $X'$ is renamed as $X$.  The above is a form of the universal coefficient
theorem which can be used to calculate $H^*(X,\M)$ from $H^*(X,\Z)$ and the
module $\M$.  The  universal coefficient theorem works for topological
cohomology where $X$ is a topological space.  The  universal coefficient
theorem also works for group cohomology where $X$ is a finite group.

Using the universal
coefficient theorem, we can rewrite \eqn{kunnZ} as
\begin{align}
\label{kunnH}
H^d(X\times X',\M) \simeq \oplus_{k=0}^d H^k[X, H^{d-k}(X',\M)] .
\end{align}
The above is valid for topological cohomology.
It is also valid for group  cohomology:
\begin{align}
\label{kunnG}
\cH^d(G\times G',\M) \simeq \oplus_{k=0}^d \cH^k[G, \cH^{d-k}(G',\M)] ,
\end{align}
provided that both $G$ and $G'$ are finite groups.

We may apply the above to the classifying
spaces of group $G$ and $G'$. Using
$B(G\times G')=BG\times BG'$, we find
\begin{align*}
H^d[B(G\times G'),\M] \simeq \oplus_{k=0}^d H^k[BG, H^{d-k}(BG',\M)] .
\end{align*}
Choosing $\M=\RZ$ and using \eqn{HdR}, we have
\begin{align}
\label{kunnGGp}
&\ \ \ \
 \cH_B^d(G\times G',\RZ)=
 H^{d+1}[B(G\times G'),\Z]
\nonumber\\
&= \oplus_{k=0}^{d+1} H^k[BG, H^{d+1-k}(BG',\Z)]
\nonumber\\
&=\cH_B^d(G,\RZ)\oplus \cH_B^d(G',\RZ)\oplus 
\nonumber\\
&\ \ \ \ \ \ \ \
\oplus_{k=1}^{d-1} H^k[BG, \cH_B^{d-k}(G',\RZ)]
\end{align}
where we have used $H^1(BG',\Z)=0$.
Using 
\begin{align}
 H^d(BG,\Z)=\cH_B^d(G,\Z),\ \ \ \
 H^d(BG,\Z_n)=\cH_B^d(G,\Z_n),
\end{align}
we can rewrite the above as
\begin{align}
\label{kunnU}
\cH^d(GG & \times SG,\RZ) = \oplus_{k=0}^{d} \cH^{k}[SG,\cH^{d-k}(GG,\RZ)]
\nonumber\\
 &= \oplus_{k=0}^{d} \cH^{k}[GG,\cH^{d-k}(SG,\RZ)]
.
\end{align}
Eqn.  \ref{kunnU} is valid for any groups $G$ and $G'$.

\section{Lyndon-Hochschild-Serre spectral sequence}
\label{LHS}

The Lyndon-Hochschild-Serre spectral sequence\cite{L4871,HS5310} allows us to
understand the structure of $\cH^d(GG\gext SG,\RZ)$ to a certain
degree. (Here $GG\gext SG$ is a group extension of $SG$ by $GG$:
$SG=(GG\gext SG)/GG$.) We find that $\cH^d(GG\gext
SG,\RZ)$, when viewed as an Abelian group, contains a chain of subgroups
\begin{align}
\{0\}=H_{d+1}
\subset H_d
\subset \cdots
\subset H_1
\subset H_0
=
 \cH^d(GG\gext SG,\RZ)
\end{align}
such that $H_k/H_{k+1}$ is a subgroup of a factor
group of $\cH^k[SG,\cH^{d-k}(GG,\RZ)]$,
\ie $\cH^k[SG,\cH^{d-k}(GG,\RZ)]$
contains a   subgroup $\Ga^k$, such that
\begin{align}
 H_k/H_{k+1} \subset \cH^k[SG,\cH^{d-k}(GG,\RZ)]/\Ga^k,\ \
k=0,\cdots,d.
\end{align}
Note that
$SG$ has a non-trivial action on $\cH^{d-k}(GG,\RZ)$ as determined by the
structure $1\to GG \to GG\gext SG \to SG \to 1$.  We also have
\begin{align}
 H_0/H_{1} &\subset \cH^0[SG,\cH^{d}(GG,\RZ)],
\nonumber\\
 H_d/H_{d+1}&=H_d = \cH^d(SG,\RZ)/\Ga^d.
\end{align}
In other words, all the elements in $\cH^d(GG\gext SG,\RZ)$ can be one-to-one labeled
by $(x_0,x_1,\cdots,x_d)$ with
\begin{align}
 x_k\in H_k/H_{k+1} \subset \cH^k[SG,\cH^{d-k}(GG,\RZ)]/\Ga^k.
\end{align}

The above discussion implies that we can also use $(m_0,m_1,\cdots,m_d)$ with
\begin{align}
 m_k\in \cH^k[SG,\cH^{d-k}(GG,\RZ)]
\end{align}
to  label all the elements in $\cH^d(G,\RZ)$. However, such a labeling scheme
may not be one-to-one, and it may happen that only some of
$(m_0,m_1,\cdots,m_d)$ correspond to  the  elements in $\cH^d(G,\RZ)$.  But, on
the other hand, for every element in $\cH^d(G,\RZ)$, we can find a
$(m_0,m_1,\cdots,m_d)$ that corresponds to it.  


\bibliography{../../bib/wencross,../../bib/all,../../bib/publst,./tmp} 

\end{document}